\documentclass[pra,twocolumn,showpacs,superscriptaddress,amsfonts,amsmath,floatfix]{revtex4-1}
\usepackage{graphicx}
\usepackage{psfrag}
\usepackage{epsfig}
\usepackage{pstricks}
\usepackage{epstopdf}

\newcommand{\Journal}[4]{#1 {\bf #2}, #3 (#4)}
\newcommand{\PRL}{Phys. Rev. Lett.}
\newcommand{\PRA}{Phys. Rev. A}
\newcommand{\PRB}{Phys. Rev. B}
\newcommand{\PRD}{Phys. Rev. D}

\newcommand{\lt}{\left(}
\newcommand{\rt}{\right)}
\newcommand{\lqu}{\left[}
\newcommand{\rqu}{\right]}
\newcommand{\lgr}{\left\{}
\newcommand{\rgr}{\right\}}

\newcommand{\be}{\begin{equation}}
\newcommand{\ee}{\end{equation}}
\newcommand{\ba}{\begin{eqnarray}}
\newcommand{\ea}{\end{eqnarray}}

\newcommand{\fr}{\frac}
\newcommand{\nn}{\nonumber}
\newcommand{\tr}{{\rm{tr}}}

\begin{document}

\title {Effect of phase noise on useful quantum correlations
in Bose Josephson junctions}
\author{G. Ferrini}\email{giulia.ferrini@grenoble.cnrs.fr}
\affiliation{Universit\'e Grenoble 1 and CNRS, Laboratoire de Physique et Mod\'elisation des
Milieux Condens\'es UMR5493, B.P. 166, 38042 Grenoble, France}
\author{D. Spehner}
\affiliation{Universit\'e Grenoble 1 and CNRS, Laboratoire de Physique et Mod\'elisation des
Milieux Condens\'es UMR5493, B.P. 166, 38042 Grenoble, France}
\affiliation{Universit\'e Grenoble 1 and CNRS, Institut Fourier  UMR5582,
B.P. 74, 38402 Saint Martin d'H\`eres, France}
\author{A. Minguzzi}
\affiliation{Universit\'e Grenoble 1 and CNRS, Laboratoire de Physique et Mod\'elisation des
Milieux Condens\'es UMR5493, B.P. 166, 38042 Grenoble, France}
\author{F.W.J. Hekking}
\affiliation{Universit\'e Grenoble 1 and CNRS, Laboratoire de Physique et Mod\'elisation des
Milieux Condens\'es UMR5493, B.P. 166, 38042 Grenoble, France}
\date{\today}


\begin{abstract}
In a two-mode  Bose Josephson junction the dynamics induced by a sudden quench of the tunnel amplitude leads to the periodic
formation of entangled states. For instance,  squeezed states  are formed at short times and macroscopic superpositions of
phase states at later times. The two modes of the junction can be viewed as the two arms of an interferometer; use of entangled states allows to perform  atom interferometry beyond the classical limit.
Decoherence due to the presence of noise   degrades the quantum correlations between the atoms, thus reducing phase sensitivity  of the interferometer.
We  consider the noise induced by stochastic fluctuations of the energies of the two modes of the junction.
 We analyze its effect on squeezed states and macroscopic superpositions and study quantitatively
the  amount of quantum correlations which can be used to enhance the phase sensitivity with respect to the classical limit. To this aim we compute the
  squeezing parameter and the quantum Fisher information during the quenched dynamics.
For moderate noise intensities we show that  these useful  quantum correlations
increase on time scales beyond the squeezing regime. This suggests
multicomponent superpositions as interesting candidates for high-precision atom interferometry.
\end{abstract}

\maketitle

\section{INTRODUCTION}

Confined ultracold atomic gases are promising candidates for implementing quantum simulators and for
applications in quantum technology, due to the high controllability of the experimental
parameters such as the atomic interactions~\cite{Julienne} and the geometry of the  trap~\cite{Lesanovsky,Fortagh}. Among the applications we cite high-sensitivity atom interferometry, which can be used for enhancing the precision in atomic clocks and in magnetic field sensors \cite{Appel,Schleier,Oberthaler10,Treutlein10}. Of particular interest are Bose Josephson junctions formed by two  modes of a Bose-Einstein condensate.  The modes may  correspond either
to two internal states of the condensed atoms in a single potential
well or to two spatially separated wavefunctions in a double well.
 During the dynamics following a sudden quench of the tunnel amplitude connecting the two modes,  squeezed states are formed at early times.
 It has been shown theoretically~\cite{Wineland,Kitagawa93, Sorensen} and
experimentally~\cite{Oberthaler10,Treutlein10} that these states
can be used to
estimate  phase shifts with  sensitivity below the shot noise limit, the limit one obtains using  classical states.
The highest possible phase sensitivity, limited by quantum uncertainty only,  can be achieved by using macroscopic superpositions of e.g. atomic phase states~\cite{Giovannetti, Smerzi09}.
Such superpositions  are however formed at later times during the quenched
dynamics of the BJJ
~\cite{varie_gatti,noi,Piazza08}.
They are expected to be very fragile with respect to decoherence effects
caused by particle losses~\cite{Sinatra}, collisions with thermal
 atoms~\cite{Anglin,Witthaut}, interaction with the electromagnetic field~\cite{Moore06}, and random fluctuations of the trapping potential~\cite{Khodorkovsky}.

In this work we  consider the effects of phase noise
 on the states formed during the quenched dynamics of the BJJ.
Phase noise  is  induced by stochastic fluctuations of the energies of the two modes of the BJJ.
Together with
atom losses, such a noise is one of the main sources of  decoherence
in the experiments of Refs.~\cite{Oberthaler08,Oberthaler10,Treutlein10}.
Interestingly, macroscopic superpositions of phase  states in BJJs  have been shown to be  robust with respect to phase noise,  their
decoherence rate being independent of the total number of atoms in the condensate  ~\cite{Ferrini_10}. Below  we show that these long-lived states
can be useful in interferometry to improve phase sensitivity.
In particular, we compare the best possible phase sensitivity obtained with the state of the BJJ at the times of formation of macroscopic superpositions to the one obtained at  earlier times
when squeezed states are produced.
This allows us to determine which are the most useful quantum states
for interferometric applications  in the presence of phase noise.
In order to quantify the amount of quantum correlations useful for interferometry,
we calculate the quantum Fisher information. 
In the theory of estimation of an unknown  parameter, this quantity is related to the bound on the precision with which the unknown parameter
- in interferometry, the phase shift - can be determined~\cite{Caves94,Giovannetti}.
We show that for moderate noise intensities the quantum Fisher information at the time of formation of the  first superpositions of phase states exceeds the one  found at the time at which squeezed states appear.
In other words, despite the action of phase noise, a  better phase sensitivity can be reached  by using states produced at times later than the time for optimal squeezing.

The paper is organized as follows. In Sec.II we recall the definition of the two parameters
relevant in interferometry, i.e. the coherent spin squeezing and the quantum Fisher
information, and link them to multiparticle entanglement. In Sec.III we present the model
which describes the quenched time  evolution of a BJJ both in the absence and in the presence of noise.  A peculiarity of the quenched time evolution is the formation of multicomponent superpositions of phase states, and we illustrate the effect of  phase noise on those states in Sec.IV by calculating a suitable probability distribution.
In Sec.V we compute the coherent spin squeezing and the quantum Fisher information during the quenched dynamics of the BJJ, first
in the absence and then in the presence of noise, and study quantitatively the loss of useful correlations as a function of time. Finally, Sec. VI contains some concluding remarks.

\section{Phase estimation in atom interferometry}
\label{secII}

\subsection{Phase estimation and the quantum Fisher information}
\label{sec-def-Fisher}

\begin{figure}
\begin{minipage}{.70\columnwidth}
\includegraphics*[width=\columnwidth]{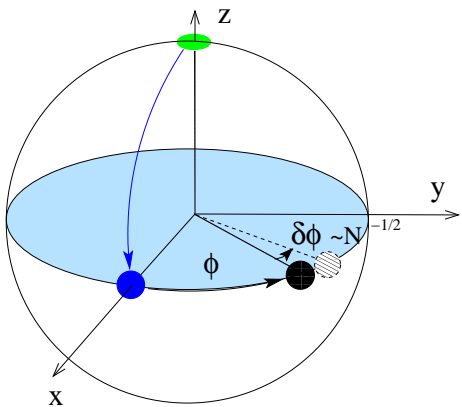}
\end{minipage}
\caption{(Color online)
Rotations on the Bloch sphere in the interferometric scheme: the input coherent state  at the north pole (green disk) is  rotated around the $y$-axis by an angle $\pi/2$ (blue disk) and afterwards around the $z$-axis by the unkown phase $\varphi$ (black disk). The precision $\Delta \varphi$ on the estimation of $\varphi$ is larger than the size $\sqrt{N}/2$ of the disk,   representing the angular momentum fluctuations, divided by the radius $N/2$ of the sphere. The last rotation around the $y$-axis is not represented.  }
\label{fig:Bloch_sphere}
\end{figure}

 The goal in interferometry is to
estimate an unknown phase shift $\varphi$ with the highest
possible precision. In atom interferometry, an input state is first transformed into a
superposition of two modes, analogous to the two arms of an optical interferometer. These
modes acquire distinct phases $\varphi_1$ and $\varphi_2$ during the subsequent quantum
evolution. They are are finally recombined to read out interference fringes, from which
the phase difference $\varphi=\varphi_1-\varphi_2$ is inferred.
The interferometric sequence can be described by means of rotation
matrices acting on the
 two-mode vector state, that is,  by SU(2) rotation matrices
in the Schwinger representation~\cite{Kim,Yurke86}.
The generators of the rotations are the angular-momentum operators
 $\hat J_x$, $\hat J_y$, and $\hat J_z$, related to the annihilation operator
$\hat{a}_j$ of an atom in the mode
$j=1,2$ by $\hat J_x=(\hat a^\dagger_1 \hat a_2+ \hat a^\dagger_2 \hat a_1)/2$,
$\hat J_y=-i (\hat a^\dagger_1 \hat a_2-\hat a^\dagger_2 \hat a_1)/2$, and
$\hat J_z  \equiv \hat n = (\hat a^\dagger_1 \hat a_1- \hat a^\dagger_2 \hat a_2)/2$,
the latter being the number imbalance operator.
Let us consider the case where the two modes correspond to two internal states 
of the atoms in an optically trapped Bose-Einstein condensate.
The total number $N$ of atoms in the condensate is assumed to be fixed and
all atoms are initially in the lower energy  state (mode $j=1$).
The input state is then $|n_z=N/2 \rangle$, where
$|n_z =n\rangle \equiv | n_1,n_2\rangle$ denotes the Fock state satisfying
$\hat J_z |n_z=n\rangle = n | n_z=n\rangle$,
$n_1=N/2+n$ and $n_2 =N/2-n$ being the number of
atoms in the lower and upper modes, respectively.
The application of a $\pi/2$ laser pulse  with frequency
in resonance with the two internal levels plays the role of a beam splitter in optical
interferometers.
It brings the input state onto the coherent state $| \theta=\pi/2,\phi=0\rangle$, where the
SU(2) coherent states are defined as ~\cite{husimi}
\begin{equation} \label{eq-coherent_state}
|\theta, \phi \rangle=\sum_{n_1=0}^{N}
\left(  \begin{array} {c} N \\ n_1 \end{array}\right)^{1/2}
\frac{\alpha^{n_1}}{(1+|\alpha|^2)^{N/2}}\, |n_1,N-n_1\rangle
\end{equation}
with $\alpha =\tan(\theta/2) \exp(-i\phi)$. It is easy to show that
$|\theta, \phi \rangle
\propto (e^{-i \phi} \sin (\theta/2) a_1^\dagger + \cos (\theta/2) a_2^\dagger )^N | 0 \rangle$
(here $| 0\rangle$ is the vacuum state), meaning that all atoms occupy the same one-atom state.
A coherent state can be visualized as a  disc of diameter
$\sqrt{N}/2$ on the Bloch sphere of
radius $N/2$, centered at $N (\sin \theta \cos \phi ,\sin \theta \sin \phi , - \cos \theta)/2$.
The coordinates of the center are the expectation values of
the angular momentum operators $\hat{J}_x$,  $\hat{J}_y$, and $\hat{J}_z$ in
$|\theta, \phi \rangle$, whereas the diameter of the disc
gives the quantum fluctuations of $\hat{J}_{\vec{n}}= \vec{{J}} \cdot \vec{n}$
in the directions $\vec{n}$  tangential to the sphere.
The coherent states with $\theta = \pi/2$ on the equator of the Bloch
sphere are referred to as \emph{phase states}.
The Fock state $|n_z=N/2 \rangle$ is a coherent state with $\theta=\pi$, located at the  north pole
of the Bloch sphere. The action of the beam splitter is therefore  a rotation of the atomic  state around
the  $y$-axis by an angle of $\pi/2$ radians, leading to the phase state
$| \theta=\pi/2,\phi=0\rangle$. Then the  state is rotated around the $z$-axis by the
free evolution, the phase accumulation being due to a different energy shift between the two states.
This rotation is the analog of the different phase paths in the two arms of an optical
interferometer.
The consecutive rotations of the input state on the Bloch sphere are represented in
Fig.\ref{fig:Bloch_sphere}.
Finally, by recombining the two paths, the state is rotated again around the $y$-axis by
an angle of $-\pi/2$ radians.
The interferometric sequence can thus be described by a succession of three rotations,
and the output state of the linear interferometer is
\be
\label{eq:curva_varying_theta}
| \psi_{\text{out}}  \rangle
= e^{-i \fr{\pi}{2} \hat J_{y} }e^{-i\varphi \hat J_{z} }
e^{i \fr{\pi}{2} \hat J_{y} } | \psi_{\text{in}}  \rangle
= e^{-i \varphi \hat J_{x}}  | \psi_{\text{in}} \rangle,
\ee
where $| \psi_{\text{in}} \rangle$ is the input state, assumed here to be pure.
More generally, the output state of the interferometer is
\be
\label{eq:interferometer}
\hat{\rho}_{\text{out}}(\varphi) = e^{-i \varphi \hat J_{\vec{n}}} \hat{\rho}_{\text{in}}
e^{i \varphi \hat J_{\vec{n}}}.
\ee
where $\hat{\rho}_{\text{in}}$ is the input density matrix and $\vec{n}$ the unit vector representing the effective rotation axis associated to a given interferometric sequence.

In a typical experiment one has access to the probability distribution associated to the operator  $\hat J_z$ measured with respect to the output state.
This quantum distribution depends
on the phase shift $\varphi$. The latter is then
determined by means of a statistical estimator
depending on the results of the measurements of
$\hat J_z$ in the output state.
The precision $\Delta \varphi$
with which the phase shift $\varphi$ can be determined
depends on the chosen estimator, on the input state and on the measurement performed on the output state. Optimizing over all possible measurements, 
the best precision that can be achieved
for a given input state $\hat{\rho}_{\text{in}}$
is, according to the Cram\'er-Rao bound~\cite{Caves94},
\be
\label{eq:Cramer_Rao}
\Delta \varphi \geq (\Delta \varphi)_{\text{best}}
= \fr{1}{\sqrt{m} \sqrt{F_Q \lqu \hat{\rho}_{\text{in}}, \hat{J}_{\vec{n}} \rqu}},
\ee
where $m$ is the number of measurements and
$F_Q [ \hat{\rho}_{\text{in}},\hat{J}_{\vec{n}} ]$ the quantum Fisher information
given by~\cite{Caves94}
\be
\label{eq:Fisher_computing}
F_Q \lqu \hat{\rho}_{\text{in}},\hat{J}_{\vec{n}} \rqu
=  2 \sum_{l,m,p_l+p_m>0} \fr{(p_l - p_m)^2}{p_l + p_m} |\langle l |\hat J_{\vec{n}} | m \rangle | ^2 \, ,
\ee
$\{ | l \rangle \}$ being an orthonormal basis diagonalizing
$\hat{\rho}_{\text{in}} = \sum_l p_l | l \rangle \langle l | $
(with $p_l \geq 0$ and $ \sum_l p_l = 1$).
The Fisher information (\ref{eq:Fisher_computing})
depends on the input state and on the direction $\vec{n}$ of the interferometer.
It has the meaning of the square of a "statistical speed" at which the state evolves
along the curve defined by Eq.(\ref{eq:interferometer}) in the space of density matrices
when the parameter $\varphi$ is varied~\cite{Caves94,Smerzi09}: if one
increases $\varphi$ starting from $\varphi=0$ with a fixed velocity $\dot{\varphi}$, the faster the state
(\ref{eq:interferometer}) becomes distinguishable from $\hat{\rho}_{\text{in}}$, the larger
is its quantum Fisher information $F_Q$.
Hence the bound (\ref{eq:Fisher_computing}) relates the problem of estimating a
phase shift in an interferometer to the problem of distinguishing neighbouring quantum
states \cite{Caves94}. Indeed, the quantum Fisher information is
related to the Bures riemannian distance on the space of density matrices~\cite{Uhlmann76}.

For pure input states $| \psi_{\text{in}} \rangle$, the quantum Fisher information is
given by the quantum
fluctuation
$(\Delta {J}_{\vec{n}})^2 = \langle \psi_{\text{in}} |  \hat{J}_{\vec{n}}^2 | \psi_{\text{in}} \rangle -
\langle \psi_{\text{in}} |  \hat{J}_{\vec{n}} | \psi_{\text{in}} \rangle^2$
of $\hat{J}_{\vec{n}}$,
\be
\label{eq:fisherpure}
F_Q \lqu |\psi_{\text{in}}\rangle , \hat{J}_{\vec{n}} \rqu = 4 ( \Delta {J}_{\vec{n}})^2\,.
\ee
This allows to reinterpret the Cram\'er-Rao lower bound (\ref{eq:Cramer_Rao})
as a generalized uncertainty principle
\be \label{eq-uncertainty_principle}
\Delta \varphi \,\Delta {J}_{\vec{n}} \geq \fr{1}{2 \sqrt{m}}\,,
\ee
in which the generator $\hat{J}_{\vec{n}}$  of the transformation (\ref{eq:interferometer})
and the phase shift $\varphi$ play the role of two conjugate variables
- $\varphi$ being here not an observable but a parameter \cite{Caves94}.
For instance, the
Fisher information of a phase state $|\psi_{\text{in}}\rangle = | \theta=\pi/2, \phi \rangle$
in the directions
$\vec{n} = \vec{e}_x$, $\vec{e}_y$, and $\vec{e}_z$
are equal to $N \sin^2 \phi$, $N \cos^2 \phi$, and $N$, respectively.
According to (\ref{eq-uncertainty_principle}),
for such a state the best precision that can be achieved on the phase shift is
$(\Delta \varphi)_{\rm{best}}= 1/\sqrt{N m}\equiv(\Delta \varphi)_{{SN}}$, corresponding to the
{\it shot-noise limit} of independent atoms.

The saturation of the bound (\ref{eq:Cramer_Rao}) requires both a suitable
classical post-processing on the $m$ outcomes of the measurements (e.g. the maximum
likelihood estimation in the limit of large $m$ \cite{Caves94}) and the knowledge
of the optimum observable to measure. This latter task can be difficult as the optimum
measurement may depend on the phase shift itself \cite{Caves96}.

It can be shown~\cite{Giovannetti, Smerzi09} that for any separable input state
$\hat{\rho}_{\text{in}}$,  $F_Q [ \hat{\rho}_{\text{in}},\hat{J}_{\vec{n}} ] \leq N$, so that \be
\label{eq:fisher_criterion}
F_Q \lqu  \hat{\rho}_{\text{in}},\hat{J}_{\vec{n}} \rqu > N
\ee
is a sufficient condition for $ \hat{\rho}_{\text{in}}$ to be entangled; in other words, $F_Q-N$ is an entanglement witness.
By Eq.(\ref{eq:Cramer_Rao}), the inequality (\ref{eq:fisher_criterion}) is
a necessary and sufficient condition for sub-shot noise
sensitivity
$(\Delta \varphi)_{\text{best}}  < (\Delta \varphi)_{SN}$.
In what follows, the input states leading to such a condition are called
 \emph{useful states} for interferometry (or, more briefly, ``useful states'').
It is worthwhile to stress that the inequality (\ref{eq:fisher_criterion}) is
not a necessary condition for entanglement: indeed,
there exists entangled states  which are not useful for
interferometry, that is, with a Fisher information $F_Q \leq N$~\cite{Smerzi09,Smerzi10}.

The quantum Fisher information is bounded by  $N^2$. This is easy to show for pure states by noticing that the
largest square fluctuation of $\hat{J}_{\vec{n}}$ in Eq.(\ref{eq:fisherpure}) is smaller or equal to
$N^2/4$ (see~\cite{Giovannetti}); for mixed states this follows from the convexity of
$F_Q$ (see~\cite{Smerzi09}). According to Eq.(\ref{eq:Cramer_Rao}),
 the best sensitivity that  can be achieved in linear interferometers~\cite{Giovannetti2}
is then
$(\Delta \varphi)_{\text{best}} = (\Delta \varphi)_{HL} \equiv 1/N$.
This corresponds to the so-called \emph {Heisenberg limit}.
This limit is reached using highly entangled atoms as
 input state, e.g. the macroscopic superposition given by the
 so-called NOON state
$| \psi_{\text{NOON}} \rangle = (|N,0 \rangle + e^{i \alpha} |0,N \rangle)/\sqrt{2}$, with
$\alpha$ being a real phase.
The quantum Fisher information of a NOON state is equal to $N^2$ in the direction
$\vec{n}=\vec{e}_z$. It is instructive to compare this result with the value of the quantum Fisher information for a statistical mixture of the same states,
$\hat{\rho}_{\text{NONO}} = (|N,0 \rangle \langle N,0| + |0,N \rangle \langle 0,N|)/2$.
The latter is
found with the help of Eq.(\ref{eq:Fisher_computing}) to be equal to $N$ in all
directions $\vec{n}$ in the $(xOy)$-plane and to vanish in the direction $\vec{e}_z$.
Therefore, the scaling of $F_Q$ like $N^2$ for $\hat{\rho}_{\rm NOON} =
| \psi_{\text{NOON}} \rangle \langle \psi_{\text{NOON}}|$ is
due to the presence  of the off-diagonal terms
 $\hat{\rho}_{\rm NOON} - \hat{\rho}_{\text{NONO}} =
(e^{-i \alpha} |N,0 \rangle \langle 0, N| + e^{i \alpha} |0,N \rangle \langle N, 0|)/2$.

To summarize, the study of the quantum Fisher information and  its scaling with the number
of atoms allows to quantify
the amount of quantum correlations which can be used to enhance the precision on
the phase shift in interferometry.

\subsection{Coherent spin squeezing}

Atomic squeezed states are examples of nonclassical states useful for interferometry,  which have been recently realized experimentally \cite{Appel,Schleier,Oberthaler10,Treutlein10}.
The coherent spin squeezing parameter quantifies the angular-momentum fluctuations along the direction $\vec{n}$ \cite{Wineland, Sorensen} according to
\be
\label{eq:squeezing_n}
\xi_{\vec{n}}^2 \lqu \hat{\rho}_{\text{in}},\hat{J}_{\vec{n}} \rqu =  \fr{N \Delta^2 \hat{J}_{\vec{n}} }{\langle \hat{J}_{\vec{p}_1} \rangle^2 + \langle \hat{J}_{\vec{p}_2} \rangle^2},
\ee
where
\ba
\label{eq:vector_perp}
\vec{p}_1 &=& \cos \phi\, \vec{e}_x + \sin \phi \, \vec{e}_y \nn \\
\vec{p}_2 &=& - \cos \theta \sin \phi \,\vec{e}_x + \cos \theta \cos \phi\, \vec{e}_y
+ \sin \theta \,\vec{e}_z
\ea
are unit vectors perpendicular to
\be
\label{eq:vector_generic}
\vec{n} = \sin \theta \sin \phi \,\vec{e}_x - \sin \theta \cos \phi \,\vec{e}_y
+ \cos \theta \, \vec{e}_z,
\ee
and $\langle \cdot \rangle = \tr ( \cdot \hat{\rho}_{\text{in}} )$ is the mean expectation in state
$\hat{\rho}_{\text{in}}$.
A state $\hat{\rho}_{\text{in}}$ is said to be squeezed in the direction $\vec{n}$ if the squeezing parameter satisfies
\be
\label{eq:squeezing_criterion}
\xi_{\vec{n}}^2 \lqu \hat{\rho}_{\text{in}},\hat{J}_{\vec{n}}  \rqu < 1.
\ee
It is known that Eq.(\ref{eq:squeezing_criterion}) provides both a sufficient
(but not necessary) condition for sub-shot noise sensitivity \cite{Wineland} and a
sufficient (but not necessary) condition for entanglement of
$\hat{\rho}_{\text{in}}$ \cite{Sorensen}. We remark that the squeezing criterion (\ref{eq:squeezing_criterion})
does not recognize  all useful states in interferometry. For instance,
  the NOON state does not fulfill this criterium even though it leads to the best
achievable precision.
The criteria for entanglement and sub-shot noise sensitivity are summarized in Table I.

\begin{table}
\begin{tabular}{|c|c|}
\hline
Phase estimation & Entanglement \\
\hline
 $F_Q \lqu \hat{\rho}_{\text{in}} \rqu > N \Leftrightarrow  (\Delta \varphi)_{\text{best}} <
(\Delta \varphi)_{SN}$ &  $F_Q \lqu \hat{\rho}_{\text{in}} \rqu > N \Rightarrow  \hat{\rho}_{\text{in}} \neq \hat{\rho}_{\text{sep}}$ \\
\hline
$\xi^2 \lqu \hat{\rho}_{\text{in}}\rqu < 1 \Rightarrow  (\Delta \varphi)_{\text{best}}
< (\Delta \varphi)_{SN}$ &  $\xi^2 \lqu \hat{\rho}_{\text{in}}\rqu < 1 \Rightarrow
\hat{\rho}_{\text{in}} \neq \hat{\rho}_{\text{sep}}$ \\
\hline
\end{tabular}
\label{tab:tab1}
\caption{Necessary and/or  sufficient conditions for sub-shot noise
phase sensitivity in an atom interferometer and multiparticle entanglement in terms
of the quantum Fisher information and spin-squeezing parameter.}
\end{table}

\subsection{Optimum coherent spin squeezing and quantum Fisher information}

The quantum Fisher information $F_Q$ and the spin squeezing parameter $\xi$
introduced in the previous subsections  depend on the direction of
the generator which defines the interferometric sequence (\ref{eq:interferometer}).  For instance, as shown in Sec.\ref{sec-def-Fisher}, $F_Q [ | \psi_{\text{NOON}} \rangle,\hat{J}_z ] =N^2$, corresponding to a maximally entangled state, whereas in the perpendicular directions
 $F_Q [ | \psi_{\text{NOON}} \rangle,\hat{J}_x ] = F_Q [ | \psi_{\text{NOON}} \rangle,\hat{J}_y ] =N$.
Hence, in order  to
quantify the useful correlations of a quantum state, one needs to optimize
$F_Q$ and $\xi$ over all the possible directions
by defining~\cite{Smerzi10}
\be
\label{eq:squeezing}
\xi^2 \lqu \hat{\rho}_{\text{in}} \rqu
\equiv \min_{\vec{n}} \,\xi^2_{\vec{n}} \lqu \hat{\rho}_{\text{in}} \rqu
\; , \;
F_Q \lqu \hat{\rho}_{\text{in}} \rqu \equiv \max_{\vec{n}}
F_Q \lqu \hat{\rho}_{\text{in}},\hat{J}_{\vec{n}} \rqu \,.
\ee
Let us consider the  $3\times 3$ real symmetric covariance  matrix $\gamma [ \hat{\rho}_{\text{in}} ]$
with matrix elements
\be
\label{eq:covariance_general}
\gamma_{ij} \lqu \hat{\rho}_{\text{in}} \rqu
 = \fr{1}{2} \sum_{l,m,p_l+p_m>0} \frac{(p_l-p_m)^2}{p_l+p_m}
\Re \mbox{e } \left[\langle l| \hat{J}_i|m \rangle  \langle m| \hat{J}_j|l \rangle \right]
\ee
where $\{ | l\rangle \}$ is
the orthonormal eigenbasis of $\hat{\rho}_{\text{in}}$ as in Eq.(\ref{eq:Fisher_computing}).
According to standard linear algebra, the maximum of
$F_Q [ \hat{\rho}_{\text{in}},\hat{J}_{\vec{n}} ]
= 4 ( \vec{n} , \gamma\lqu \hat{\rho}_{\text{in}} \rqu \vec{n} )$ over all
unit vectors $\vec{n}$ is equal to
\be
\label{eq:def_fisher_2}
F_Q \lqu \hat{\rho}_{\text{in}} \rqu  =
4 \gamma_{\rm max}\,,
\ee
$\gamma_{\rm max}$ being the largest eigenvalue of
the matrix $\gamma [ \hat{\rho}_{\text{in}} ]$.
In the following it will be useful to define also the matrix
\be
\label{eq:def_matrix_gamma_0}
 G_{ij}[\hat \rho] \equiv  \fr{1}{2} \langle \hat{J}_{i} \hat{J}_{j}
+ \hat{J}_{j} \hat{J}_{i} \rangle - \langle \hat{J}_{i} \rangle \langle \hat{J}_{j} \rangle,
\ee
where $\langle \ldots \rangle=\tr(\dots \hat \rho)$, with $\hat \rho$ being the system density matrix.
Note that for pure input states $|\psi_{\text{in}}\rangle$ the matrix  $\gamma_{ij} [|\psi_{\text{in}}\rangle]$ reduces to the matrix  $G_{ij}[|\psi_{\text{in}}\rangle\langle \psi_{\text{in}}|]$, which is easier to compute than the more general expression
(\ref{eq:covariance_general}).
The optimum quantum Fisher information is then
given (up to a factor four)  by the largest uncertainty
of the angular momentum operators $\hat{J}_{\vec{n}}$ (see Eq.(\ref{eq:fisherpure})).
For the sake of brevity, in the following we will omit both the adjective "optimum" and the explicit dependence on the input state, designating the optimum coherent spin squeezing and the optimum quantum Fisher information respectively by $\xi^2$ and $F_Q$, unless where source of confusion.

\section{Quenched dynamics of a BJJ}
\label{sec:3}
\subsection{Noiseless dynamics}

We describe a Bose Josephson junction (BJJ) by  a two-mode Hamiltonian~\cite{Milburn}, which in
terms of the angular-momentum operators introduced in Sec.\ref{secII} reads
\begin{equation}
\label{eq:ham_dopo_mapping}
\hat{H}^{(0)}=\chi \hat J_z^2 -\lambda \hat J_z -2 K \hat J_x\;.
\end{equation}
This Hamiltonian  models both
 a single-component Bose gas trapped in a double-well potential~\cite{Oberthaler08}
- external BJJ -
and  a binary mixture of atoms
in distinct hyperfine states trapped in a single well~\cite{Hall98,Oberthaler10}
- internal BJJ. In the external BJJ the two modes $i$
correspond to the lowest-energy
spatial modes in each well.
For the internal BJJ, the two relevant modes are the two hyperfine states.
The first term in the Hamiltonian (\ref{eq:ham_dopo_mapping}) describes the
repulsive atom-atom interactions; for the external BJJ,
$\chi$ is half the sum of the interaction energies
$U_i$ in the two modes,
whereas for the internal BJJ
$\chi=(U_1+U_2)/2 - U_{12}$ also depends on the
inter-species interaction $U_{12}$. In both cases,
$\lambda=\Delta E +( N-1) (U_2-U_1)/2$ is related
to the difference $\Delta E = E_2 - E_1$
between the energies of the two modes.
The last term  in (\ref{eq:ham_dopo_mapping}) corresponds to tunnelling
between the two wells or, for the internal BJJ, to a resonant laser field
coupling the two hyperfine states,  which can serve to implement a 50\% beam splitter
as described in Sec.\ref{sec-def-Fisher}.
Both $\chi$ and $K$ are experimentally tunable parameters.
In internal BJJs arbitrary rotations of the
form (\ref{eq:interferometer}) can be performed with a suitable combination of laser
pulses. Such rotations are typically realized fast
enough to neglect the non-linear effects induced by the interactions \cite{Oberthaler10}.
The residual effect of interactions on the interferometric
sequence has been recently addressed in Refs.~\cite{Augusto_schmiedmayer,Vardi2010}.

We consider the dynamical evolution induced by a sudden quench of the tunnel amplitude $K$ to
 zero, taking as initial state the phase state $|\theta=\pi/2,\phi=0\rangle$.
This is
 the ground state of the Hamiltonian (\ref{eq:ham_dopo_mapping})
in the regime  $K N\gg \chi$ where tunelling dominates interactions
(in the internal BJJ this state can be produced by using a laser pulse
as explained in Sec.\ref{sec-def-Fisher}).
We assume a fixed total number of atoms $N$, that is, we do not account for atom losses.
Going to the rotating frame \cite{Gross_thesis},
 we may suppose that $\lambda=0$.
In the absence of noise, the atomic state
\be
\label{eq:time_ev_noisless}
|\psi^{(0)} (t)\rangle = e^{-i\chi \hat{J}_z^2 t} |\theta=\pi/2,\phi=0\rangle
\ee
displays a periodic evolution with period
$T=2 \pi/\chi$ if $N$ is even and $T/2$ if $N$ is odd,
corresponding to  the revival time. At intermediate times the dynamics
drives the system first into a squeezed state at short times, then at times
$t_q=\pi/(\chi q)$ to a  macroscopic superposition of $q$ phase states
given by~\cite{varie_gatti,noi,Piazza08}
\be
\label{eq-q-component_cat_state}
| \psi^{(0)} (t_q) \rangle
 = u_0 \sum_{k=0}^{q-1} c_{k,q} \Bigl| \frac{\pi}{2}, \phi_{k,q} \Bigr\rangle
\ee
where   $|u_0|=q^{-1/2}$, $\phi_{k,q}= (2k-N)\frac{\pi}{q}$,
$c_{k,q} = e^{i \pi k^2/q}$  if $q$ is even, and $\phi_{k,q}=(2 k+1-N)\frac{\pi}{q}$,
$c_{k,q} = e^{i \pi k (k+1)/q}$ if $q$ is odd. This follows from
Eqs.(\ref{eq-coherent_state}) and (\ref{eq:time_ev_noisless}) and the use of the  Fourier expansion
$e^{-i \pi n^2_1 /q} = u_0 \sum_{k=0}^{q-1} e^{i \pi k^2/q} e^{-2i \pi n_1 k/q}$.
As in the case of the NOON state discussed in Sec.\ref{sec-def-Fisher},
the two-component
superposition formed at $t=t_2=T/4$ leads to the best achievable phase sensitivity if used as an
input state of the interferometer described by Eq.(\ref{eq:interferometer})~\cite{Smerzi09}.
We will show in
 Sec.\ref{secII} that the multicomponent superpositions with $3 \leq q\lesssim \sqrt{N}$,
which are formed at
earlier times $t_q < t_2$, lead to comparable phase sensitivities
up to a factor of two.
The fact that squeezed states and macroscopic superpositions of phase states
 are intrinsically produced
 by   interatomic interactions
yields a major advantage of atomic interferometers over optical ones.

\subsection{Dynamics of a noisy BJJ}
\label{sec-noisy_BJJ}

The presence of noise during the dynamical evolution of the BJJ
affects the preparation of the aforementioned
useful entangled states \cite{nota3}.
We focus here on phase noise caused by a randomly fluctuating energy difference
$\Delta E(t)$
between the two modes, assuming  that the
interaction energies $U_1$ and $U_2$ are not fluctuating.
In the single-well
experiment~\cite{Oberthaler10} (internal BJJ), such a noise is
induced by fluctuations of the
magnetic field, whereas in the double-well experiment~\cite{Oberthaler08} (external BJJ)
it is induced by fluctuations in direction
of the laser beam producing the double-well
potential with respect to the trapping potential.
The corresponding time evolution which follows the sudden quench  $K\to 0$ is described
by the time-dependent Hamiltonian
 \begin{equation}
\label{eq:hamiltonian_noise}
\hat{H}(t) = \chi \hat{J_z}^2- \lambda(t) \hat J_z.
\end{equation}
 Even in presence of noise, the time evolution following the quench can be exactly
integrated since   the
noise term $\lambda(t) \hat J_z$ commutes with the noiseless Hamiltonian $\chi \hat{J_z}^2$
\cite{Ferrini_10}. For a given realization of the stochastic process $\lambda(t)$,  the state of the atoms at time $t$ is
 \be
|\psi (t) \rangle = e^{-i \phi(t)  \hat{J}_z }|\psi^{(0)} (t) \rangle
\ee
 where  $\phi (t) \equiv - \int_0^{t}{  d\tau \lambda(\tau)}$ and
$|\psi^{(0)} (t) \rangle$
is the time-evolved state (\ref{eq:time_ev_noisless}) in the absence of noise.
The system density matrix is then obtained by
$\hat{\rho}(t) = \overline{| \psi (t)\rangle \langle \psi (t) |}
= \int d P \lqu \lambda \rqu |\psi (t) \rangle \langle \psi (t)|$, where
the overline denotes the average over the noise realizations.
The introduction of the distribution probability for the random angle $\phi(t)$,
\be
\label{eq:distribution_phase}
f(\phi,t) = \int d P \lqu \lambda(t) \rqu \delta(\phi - \phi(t))\,
\ee
allows to write it as
\be
\label{eq:density_matrix_8}
\hat{\rho}(t)
 =
  \int_{- \infty}^{\infty} d\phi \, f(\phi, t) \,
   e^{-i \phi  \hat{J}_z } \hat{\rho}^{(0)}(t) e^{i \phi  \hat{J}_z } \,,
\ee
where $\hat{\rho}^{(0)}(t) = |\psi^{(0)} (t) \rangle \langle \psi^{(0)} (t) |$ is
the density matrix in the absence of noise.
Under the hypothesis of a gaussian noise (see Appendix\ref{appA}) the probability distribution (\ref{eq:distribution_phase})
reads
\be
\label{eq:fgauss}
f(\phi,t) =\fr{1}{\sqrt{2 \pi} a(t)}  e^{-\fr{(\phi +  \overline{\lambda} t)^2}{2 a^2(t)}},
\ee
where $\overline{\lambda}= \overline{\Delta E} + (N-1)(U_2-U_1)/2$ and
the variance $a^2(t)$ is given in terms of the  noise  correlation function
\be
h(\tau - \tau') = \overline{\lambda(\tau) \lambda(\tau') }-\overline{\lambda}^2
= \overline{\Delta E (\tau)\Delta E (\tau')} - \overline{\Delta E}^2
\ee
by
\be \label{eq-a2}
a^2(t) = \int_0^{t} d\tau \int_0^{t} d\tau' h(\tau - \tau')\,.
\ee
We note that $h$ depends only on the time difference $\tau - \tau'$  by the stationarity of
the stochastic process $\lambda(t)$, which also implies $\overline{\lambda(t)}= \overline{\lambda(0)}\equiv \overline{\lambda}$;
moreover, $h$ decreases to zero at sufficiently long times.
By projecting Eq.(\ref{eq:density_matrix_8}) on the Fock basis $\{ |n_z=n\rangle \}$
we obtain
\ba
\label{eq:density_matrix_9}
\hspace*{-0.2cm}
& & \langle n_z=n | \hat{\rho}(t) | n_z=n ' \rangle  =  e^{-\fr{a^2(t) (n - n')^2}{2}}
e^{i \overline{\lambda} t (n-n' )}
\nn
\\
& & \hspace*{2cm}
\times \langle n_z= n | \hat{\rho}^{(0)}(t) | n_z = n' \rangle.
\ea
In order to discuss the effect of the phase noise on the state of the atoms we briefly discuss the noise variance $a(t)$.
 We first notice that under our hypothesis, $a(t)$ and thus the decoherence factor
(given by the first exponential in the right-hand side of Eq.(\ref{eq:density_matrix_9}))
 is independent of the number
of atoms $N$ in the BJJ.
This is in contrast with the usual scenario for decoherence
which predicts stronger decoherence as the number of particles in the system is increased. As a consequence of this fact, macroscopic superpositions of the form (\ref{eq-q-component_cat_state}) are robust against phase noise, as was shown in~\cite{Ferrini_10} and will be detailed in Sec.\ref{secV} below.

Let us denote  by $t_c$ the largest time such that
$h (\tau) \simeq h (0) = \delta \lambda(0)^2 \equiv \delta \lambda^2$ and
by $T_c$ the characteristic time at which $h (\tau)$ vanishes. If the time evolution occurs
on a short scale such that $t<t_c$ then the colored nature of the noise
plays an important role and
\be
\label{eq-a2_short_time}
a^2(t)\simeq \delta \lambda^2 t^2.
\ee
If instead the time evolution occurs on a time scale much larger than the noise
correlation time $T_c$ we obtain the same result as for white noise,
\be
\label{eq-a2_Markov}
a^2(t) \simeq 2 t \int_0^\infty h(y) dy,
\ee
which corresponds to the Markov approximation.

The effect of phase noise can be partially suppressed by using the so-called spin-echo protocol  \cite{Viola_Lloyd_echo}. This strategy was followed in a recent experiment  \cite{Oberthaler10}. The analysis discussed in this section can be adapted to take into account the residual effect of phase noise when spin echo pulses are applied, see Appendix \ref{appB}.

\section{Effect of phase noise on multicomponent  macroscopic superpositions of phase states}
\label{secV}

Before analyzing the Fisher information and the spin squeezing during the quenched dynamics discussed above in detail, we wish to study the nature of the state of the atoms under phase noise at the
specific times $t_q$ which in the noiseless BJJ correspond to the formation of multicomponent superpositions of phase states. We  first illustrate the effect of the noise on the structure of the  density matrix, then we study a suitable probability distribution which is particularly sensible to
decoherence.

\subsection{ Structure of the density matrix in the Fock basis}

In the absence of noise the quenched dynamics of the Bose Josephson junction leads to the
formation of coherent superpositions with  $q$ components as given by
Eq.(\ref{eq-q-component_cat_state}).
The corresponding density matrix
$\hat{\rho}^{(0)}(t_q)=|\psi^{(0)} (t_q)\rangle\langle\psi^{(0)}(t_q)| $
has the form $\hat{\rho}^{(0)}(t_q)=\sum_{k,k'} \hat{\rho}_{k k'}^{(0)}(t_q)$, where
 the indices $k$ and $k'$ label the various components
of the superposition
 and $\hat{\rho}_{k k'}^{(0)}(t_q)=q^{-1} c_{k,q} c_{k',q}^\ast
| \pi/2,\phi_{k,q}\rangle \langle \pi/2, \phi_{k',q}|$. For general decoherence
 processes one expects that, by increasing the intensity of the noise,
$\hat{\rho}^{(0)}(t_q)$ will evolve into the statistical mixture  of phase states
$\sum_k \hat{\rho}_{k k}^{(0)}(t_q)$; moreover, the larger the atom number $N$
the weaker should be the noise strength at which this occurs \cite{Giulini,Orszag}.
It was found in \cite{Ferrini_10}
 that for the phase noise considered in Sec.\ref{sec:3}
 the actual scenario for decoherence is different from the usual one.
Indeed, the typical  noise intensity at which
the coherences between distinct phase states $| \pi/2,\phi_{k,q} \rangle$
are lost turns out to be
 independent of the atom number. This is a consequence of the fact that the decoherence factor $a(t)$ is independent of $N$, as shown in Sec.\ref{sec-noisy_BJJ}. Furthermore,  for superpositions with a large number
of components $q$, this intensity
is {\it larger than} the noise intensity at which phase relaxation occurs.
In what follows we discuss the origin of this fact.

Since the noise
is expected to destroy correlations between different components,  we decompose the density
matrix in its diagonal (intra-component) and off-diagonal (intercomponent) parts, focussing
on the latter one to quantify the decoherence. We have then
 $\hat{\rho}^{(0)}=\hat{\rho}_{\rm{d}}^{(0)}+\hat{\rho}_{\rm{od}}^{(0)}$ where
\be
\hat{\rho}_{\rm{d}}^{(0)}(t_q)=\sum_{k = 0}^{q -1} \hat{\rho}^{(0)}_{kk} (t_q)
\ee
and
\be
\hat{\rho}_{\rm{od}}^{(0)}(t_q)= \sum_{k,k' = 0; k\neq k'}^{q -1} \hat{\rho}^{(0)}_{kk'} (t_q).
\ee
Using Eq.(\ref{eq-coherent_state}) and  the identity
$\sum_{k=0}^{q-1} e^{2 i k (n'-n)\pi/q} = q$ if $n=n'$ modulo $q$ and  $0$ otherwise,
the matrix elements of $\hat{\rho}^{(0)}_{\text{d}} (t_q)$ in the Fock basis are
\ba
\label{eq:rate_noise_1}
& &
\langle n_z= n| \hat{\rho}^{(0)}_{\text{d}} (t_q) |n_z= n' \rangle
\\
\nn
& &
=\begin{cases}
\displaystyle
\fr{(-1)^{2pI(\frac{N}{2})}}{2^N}
{N \choose \fr{N}{2} + n}^{\fr{1}{2}}  {N \choose \fr{N}{2} +n'}^{\fr{1}{2}}
& \text{if $n'=n+pq$}
\\
0 & \text{if $n'\not= n$ mod $q$}
\end{cases}
\ea
%
where $p$ is an integer and $I$ denotes the integer part.
By using $\hat{\rho}^{(0)}_{\text{od}} (t_q)
= e^{- i \pi \hat{J}_z^2/q} |\theta= \pi/2,\phi=0 \rangle \langle\theta= \pi/2,\phi=0|  e^{i \pi \hat{J}_z^2/q}
- \hat{\rho}^{(0)}_{\text{d}} (t_q)$, we also get
\ba
\label{eq:matrice_off_diag2}
& &
\langle n_z=n| \hat{\rho}^{(0)}_{\text{od}} (t_q) | n_z=n' \rangle
\\
\nn
&  & =
\begin{cases}
0
& \text{if $n' = n + p q$}
\\
\displaystyle
\fr{e^{ i \alpha_{nn'}} }{2^N}
{N \choose \fr{N}{2} + n}^{\fr{1}{2}}  {N \choose \fr{N}{2} + n'}^{\fr{1}{2}}
& \text{if $n'\not= n$ mod $q$}
\end{cases}
\ea
with $\alpha_{nn'}= (n'+n-N) (n'-n) \pi/q$.
The use of Eq.(\ref{eq:density_matrix_9}) allows to obtain the corresponding expressions in
the presence of noise,
\be \label{eq:density_matrix_10}
 \langle n |  \hat{\rho}_{\text{d,od}}(t_q) | n' \rangle
 =
e^{-\fr{a^2_q (n - n')^2}{2}}  \langle n|  \hat{\rho}_{\text{d,od}}^{(0)}(t_q) | n'\rangle
\ee
up to a phase factor irrelevant for decoherence, with
$a_q \equiv a(t_q)$.
In the strong noise limit $a_q \gg 1$,  the
 off-diagonal part $\hat{\rho}_{\text{od}}$  of the atom density
matrix vanishes whereas
the diagonal part $\hat{\rho}_{\text{d}}$
tends to a matrix which is diagonal in the Fock basis,
\ba \label{eq-asymtotic_state}
\hat{\rho}_{\rm{d}} (t_q) & \rightarrow & \hat{\rho}_\infty   =
\int_{0}^{2 \pi} \frac{d \phi}{2\pi} \,|\theta = \pi/2, \phi \rangle \langle
 \theta = \pi/2, \phi |
\\
\nn
& & = \sum_{n=-N/2}^{N/2} \frac{1}{2^N} {N \choose \frac{N}{2} + n} | n_z=n\rangle \langle n_z = n|
\,.
\ea

The fact that the diagonal part of the atom density matrix decays faster than the off-diagonal part for increasing noise strengths \cite{Ferrini_10} 
is readily explained by examining the structure of the noiseless density matrices in Eqs.(\ref{eq:rate_noise_1}) and
(\ref{eq:matrice_off_diag2}). The first  off-diagonal elements of
 $\hat{\rho}_{\text{d}} (t_q)$
in the Fock basis are those for which
 $n'=n\pm q$ while the first  off-diagonal elements
of  $\hat{\rho}_{\text{od}}(t_q)$ satisfy  $n'=n\pm 1$.
Hence, it results from
Eq.(\ref{eq:density_matrix_10}) that the
off-diagonal elements of   $\hat{\rho}_{\text{d}}$  vanish at
the noise scale $a \simeq 1/q$ while  the off-diagonal elements of
$\hat{\rho}_{\text{od}}$ vanish at the larger noise scale $a \simeq 1$. In other words,
the noise is more effective in letting $\hat{\rho}_{\text{d}}$ converge to $\hat{\rho}_\infty$
 than in suppressing $\hat{\rho}_{\text{od}}$, and this effect is more pronounced the
higher is the number
of components in the superposition.
An illustration of such anomalous decoherence is given in
Fig.~2. 
The middle panels show that for
 intermediate noise strengths,  $\hat{\rho}_{\text{d}}$ has already acquired its asymptotic
diagonal form (\ref{eq-asymtotic_state}), while  $\hat{\rho}_{\text{od}}$ has not yet vanished. As we will see in Sec.\ref{sse:IV(B)} below, these results imply that,  for moderate strengths of  phase noise, macroscopic superpositions are formed and provide quantum correlations useful for interferometry.

\begin{figure}
\begin{minipage}{.98\columnwidth}
\includegraphics*[width=\columnwidth]{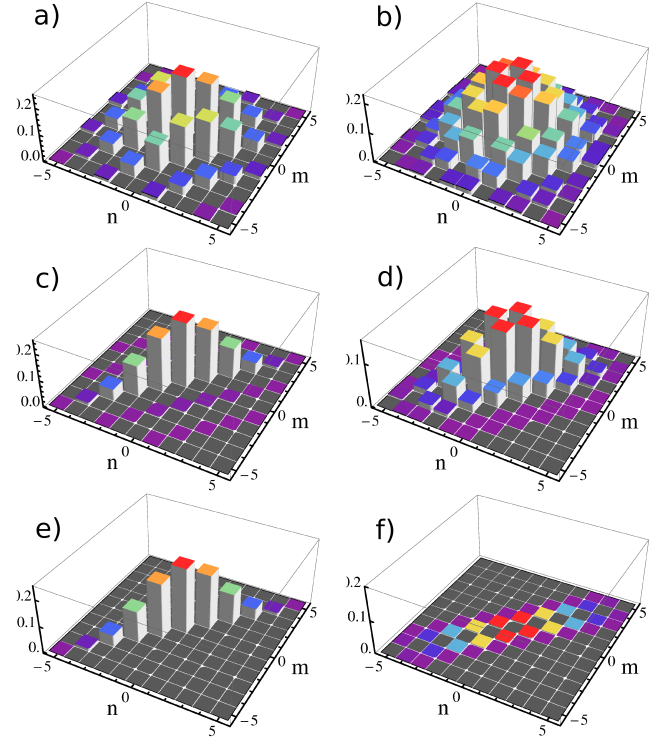}
\end{minipage}
\label{fig:density_matrix}
\caption{(Color online)  Matrix elements
of the diagonal (intracomponent) part  $\hat{\rho}_{\text{d}} (t_3)$
(panels a),c),e)) and the off-diagonal (inter-component) part $\hat{\rho}_{\text{od}} (t_3)$
(panels b),d),f)) of the  density matrix  in the Fock
basis at time $t=t_3$
 as the noise is increased from $a_3 = 0$ (a),b)) to $a_3 = 0.9$ (c),d)) and $a_3 = 2.9$ (e),f)).}
\end{figure}

\subsection{  Angular momentum distributions}

\begin{figure}
\begin{minipage}{.90\columnwidth}
\includegraphics*[width=\columnwidth]{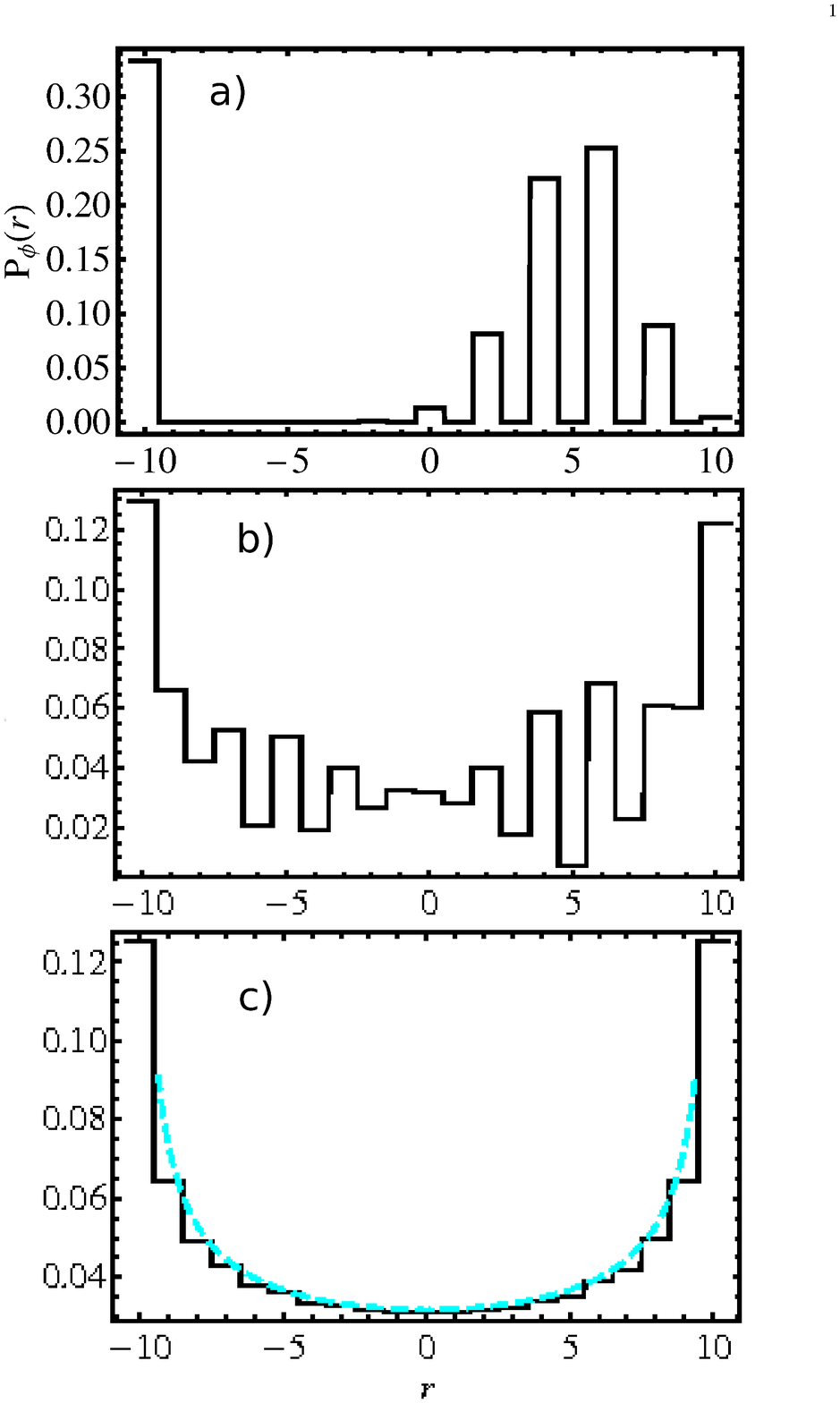}
\end{minipage}
\caption{(Color online) Probability distribution $P_{\pi/2}(r,t_3)$ of the  eigenvalues
of $\hat{J}_x$ for the three-component coherent superposition (solid lines) at
increasing noise strength from  $a_3 = 0$ (a),b)), to $a_3 = 0.9$ (c),d)) and  $a_3 = 2.9$ (e),f))
with $N=$ 20 atoms.
 The blue dashed curves indicate the large-noise intensity  and large $N$ limit
given by
Eq.(\ref{eq:proba_limiting}).}
\label{fig:probab_noise}
\end{figure}

The anomalous decoherence of the atomic state can be visualized by plotting the
probability distribution $P_{\phi}(r)$ of the eigenvalues of the
angular momentum operators
 $\hat{J}_\phi =  \hat{J}_x \sin \phi - \hat{J}_y \cos \phi $
in an arbitrary direction
of the equatorial plane of the Bloch sphere~\cite{Ferrini_2}. The presence of correlations
 among the components of the superposition  formed at time $t_q$
is indicated by interference fringes in these
distributions, which
would be  absent if  the atoms  would be in a
statistical mixture of phase states.

The probability distribution of $\hat{J}_\phi$ in the state $\hat{\rho}$
can be calculated by a straightforward generalization of the calculation
in~\cite{Ferrini_2}, as the Fourier coefficient of the characteristic function
$h_{\phi}(\eta) = \tr [e^{-i \eta \hat{J}_\phi} \hat{\rho}]$, namely,
\be
\label{eq:density_matrix_unitary_ev}
P_{\phi}(r;t) = \fr{1}{2\pi} \int_{-\pi}^{\pi} d \eta\, h_{\phi}(\eta;t) e^{i\eta r}.
\ee
For the quenched dynamics of the Bose Josephson junction
in the presence of noise, the characteristic function reads
\ba
\label{eq:h}
h_{\phi}(\eta;t) & = & \sum_{n,n'=-N/2}^{N/2} g_{n n'}(t)
\langle n_z=n |\hat{\rho}^{(0)}(t) | n_z=n' \rangle
\nn
\\
& &
\times D_{n' n}(-\phi,\eta,\phi)
\ea
where  $g_{n n'}(t) = e^{-a^2(t)(n - n')^2/2}e^{i \overline{\lambda} t (n-n' )}$
and $D_{n' n}(-\phi,\eta,\phi)$ is  the matrix element of the rotation
operator $e^{-i \eta J_\phi}$  in the Fock basis, which is given by
(see e.g.~\cite{Arecchi72}, Eq.~(D6))
\ba
&& D_{n' n}(-\phi,\eta,\phi)
= \langle n_z=n'| e^{-i \eta J_\phi} | n_z = n\rangle
\nn
\\
& &
= \sum_{k = \max \{0,n-n'\}}^{\min\{ N/2-n',N/2+n\}} (-1)^k {N \choose \fr{N}{2} + n}^{-\fr{1}{2}}{N \choose \fr{N}{2} + n'}^{-\fr{1}{2}}
\nn
\\
&& \times \fr{N!}{(\fr{N}{2} - n' - k)!(\fr{N}{2} + n - k)!k!(k + n' - n)!}
\\
&& \times \lt \sin \fr{\eta}{2} \rt^{2k + n' - n} \lt \cos \fr{\eta}{2} \rt^{N + n - n' - 2k}
e^{- i \phi(n' - n)} \,.
\nn
\ea
%
The probability distribution in the absence of noise derived in~\cite{Ferrini_2}
is recovered  by setting $ g_{n n'}(t)= 1$ in Eq.(\ref{eq:h}).

The distribution $P_{\pi/2}(r,t_3) = |\langle n_x =r |\psi^{(0)}(t_3) \rangle|^2$
of the eigenvalues of
$\hat{J}_x$ (satisfying $\hat{J}_x | n_x=r \rangle = r | n_x =r \rangle$)
is shown in Fig.\ref{fig:probab_noise} for the
three-component superposition of phase states in the absence of noise (panel a)).  Its profile displays two peaks corresponding to the
projections  on the $x$-axis of the phase states $|\theta=\pi/2,\phi=\phi_{k,3}\rangle$, $\phi_{k,3}=0$, $2\pi/3$, $4\pi/3$
(the ``phase content'' of the state, accounted for by $\hat{\rho}_{\text{d}} (t_3)$)
and interference fringes, due to the coherences between these phase states (contained in $\hat{\rho}_{\text{od}} (t_3)$).
In the presence of noise (b)-c)),
the phase profile of each component of the
superposition spreads and the characteristic peaks in the profile of the distribution
 are smeared out (phase relaxation).
At strong noise intensities, $\hat{\rho}_{\rm d} (t_q)$ approaches the steady-state given by
the density matrix (\ref{eq-asymtotic_state}), which is
 symmetric in the  $(xOy)$-plane. As a consequence,
the corresponding probability distribution $P_\phi(r,\infty) \equiv P(r,\infty)=
\tr [ \hat{\rho}_\infty | n_x = r\rangle \langle n_x = r | ]$ is independent
on $\phi$. In the semi-classical limit $N \gg 1$, this distribution can be easily calculated
since $\hat{J}_x$ takes the values $N \cos \phi/2$ in the phase states
$|\pi/2,\phi \rangle$ apart from small relative fluctuations of the order of $1/\sqrt{N}$. Hence,
recalling that $\hat{\rho}_\infty$ is a statistical mixture of the states $|\pi/2,\phi \rangle$
with equal probabilities (see Eq.(\ref{eq-asymtotic_state})),
\be
\label{eq:proba_limiting}
P(r,\infty)= c \int_0^{2\pi} \!\! d\phi \, \delta \lt \fr{N}{2} \cos \phi - r \rt = \fr{1}{\pi} \fr{1}{\sqrt{\lt \fr{N}{2} \rt^2 - r^2}}
\ee
where $c$ is a normalization factor.
The  semi-circle law (\ref{eq:proba_limiting}) is indicated by the blue dashed curve in panel c) of
Fig.\ref{fig:probab_noise}.
For finite $N$, one finds
\be
P(r,\infty)
= {N \choose \fr{N}{2} + r} \fr{1}{\pi} \fr{\Gamma \lqu \fr{N}{2} + \fr{1}{2} - r \rqu \Gamma \lqu \fr{N}{2} + \fr{1}{2} + r \rqu}{\Gamma \lqu N + 1 \rqu}. \nn
\ee
%
%
%
%

On the other hand, the vanishing of $\hat{\rho}_{\rm od} (t_q)$  tends to
diminish  the contrast of the fringes in the distribution $P_\phi (r,t_q)$, until they are completely washed
out in the asymptotic distribution (panel c) of Fig.\ref{fig:probab_noise}).
The fact that phase relaxation  occurs at a lower noise strength than  decoherence is evident
 in the panel b), where the profile of $P_{\phi}(r,t_q)$ is close to the asymptotic
distribution $P(r,\infty)$
while interference fringes due to $\hat{\rho}_{\rm od} (t_q)$ are still visible.

\section{Quantum Fisher information and coherent spin squeezing during the quenched dynamics of a BJJ}

We present in this section the estimate of the useful quantum correlations which are formed during the quenched dynamics of the Bose Josephson junction introduced in Secs.\ref{sec:3} and \ref{secV}. For this purpose, we evaluate the quantum Fisher information and the coherent spin squeezing parameter. We consider first the noiseless evolution for which analytical expressions can be obtained, then we present the numerical results for the dynamical evolution in the presence of noise.

\subsection{Dynamics in the absence of noise}
\label{sss:IV(A)}

In the absence of noise the atoms are in a pure state $| \psi^{(0)} (t)\rangle$
during all the dynamical evolution.
The covariance matrix $\gamma^{(0)}(t)$ associated to this state
is thus given by Eq.(\ref{eq:def_matrix_gamma_0}).
For the quenched dynamics described in Sec.\ref{sec:3}
by using  Eqs.~(\ref{eq-coherent_state}), (\ref{eq:time_ev_noisless}),
one finds that
\be
\label{eq:averagesJy}
\langle \hat{J}_y \rangle_t^{(0)}   =
\langle \hat{J}_z (t) \rangle_t^{(0)}  =0,
\ee
\be
\label{eq-visibiblity}
\langle \hat{J}_x \rangle_t^{(0)}  = \frac N 2 \cos^{N-1} \Bigl( \frac{2\pi t}{T} \Bigr)\equiv \frac N 2 \nu^{(0)} (t),
\ee
where $\langle ..  \rangle_t^{(0)}=\tr(... \hat\rho^{(0)}(t))$  and
$\nu^{(0)} (t)$ corresponds to the visibility  of the Ramsey fringes~\cite{Wineland}.
The  angular-momenta covariance matrix (\ref{eq:def_matrix_gamma_0}) finally reads
\begin{widetext}
\be
\label{eq:cov_matrix_unitary_t_dep}
\hspace{-1.9cm} \gamma^{(0)} (\tau)=
\left(
\begin{array}{cccccccc}
\gamma^{(0)}_x(\tau)  &  0 & \hspace{0.3cm} 0\\
\vspace{0.2cm} \\
0 &  -\fr{N}{8} \lqu (N - 1) \cos^{N-2}\lt 2 \tau \rt - (N +1) \rqu  & \hspace{0.3cm} \fr{N(N -1)}{4}  \cos^{N-2}\lt \tau \rt \sin\lt \tau \rt \\
\vspace{0.2cm} \\
0 &  \fr{N(N -1)}{4}  \cos^{N-2}\lt \tau \rt \sin\lt \tau \rt & \hspace{0.3cm} \fr{N}{4} \\
\end{array}
\right)
\ee
where we have introduced  the rescaled time $\tau=2\pi t/T=\chi t$ and
\be
\label{eq-eigenvalue_x}
\gamma^{(0)}_x(\tau) \equiv \langle (\Delta \hat{J}_x )^2\rangle_\tau^{(0)}
= \fr{N}{8} \lqu (N - 1) \cos^{N-2}\lt 2 \tau \rt + (N +1) - 2 N \cos^{2(N-1)}\lt \tau \rt \rqu \,.
\ee
The two other eigenvalues of the matrix (\ref{eq:cov_matrix_unitary_t_dep}) are
\be
\label{eq:lambdas_sol_sec}
\gamma^{(0)}_{\pm}(\tau)
= \fr{N}{16} \lqu -(N -1) \cos^{N-2}\lt 2 \tau \rt + (N +3) \pm (N -1)\sqrt{( \cos^{N-2}\lt 2 \tau \rt - 1)^2 + 16 \cos^{2(N-2)}\lt \tau \rt \sin^2 \lt \tau \rt} \rqu
\ee
\end{widetext}
(see also Ref.\cite{Devi_Sanders}).
We remark that the matrix (\ref{eq:cov_matrix_unitary_t_dep})
has the property that
its eigenvalues at times $\tau$ and $\pi - \tau$ (and, similarly, at $2\pi-\tau$) coincide, hence it suffices to discuss its behaviour at times
 $t$ belonging to the interval
$[0, T/4]$ (i.e., $\tau \in [0, \pi/2]$).

According to
Eq.(\ref{eq:def_fisher_2}), the quantum Fisher information is given by the largest eigenvalue,
\be \label{eq-Fisher_info_no_noise}
F_Q (\tau) = 4 \max \bigl\{ \gamma_x^{(0)} (\tau)\, ,\,\gamma_+^{(0)} (\tau) \bigr\}
\,.
\ee
 We demonstrate in Appendix~\ref{appC}
that the coherent spin squeezing (\ref{eq:squeezing_n}) is always optimum along a direction
contained in the $(y O z)$-plane. The optimal
spin squeezing parameter (\ref{eq:squeezing}) is
thus related to the lowest eigenvalue $\gamma^{(0)}_{-}(\tau)$ of the
submatrix ${\gamma^{(0)}}'(\tau)$ obtained by removing the first lign
and column in the matrix (\ref{eq:cov_matrix_unitary_t_dep}).
Using  Eqs.(\ref{eq:averagesJy})  and (\ref{eq-visibiblity}), one gets
\be
\label{eq:squeezing_like_davi2}
{\xi^{(0)}}^2 (\tau) = \fr{4 \gamma^{(0)}_{-}(\tau)}{N {\nu^{(0)}}^2(\tau)}
\ee
The direction of optimum squeezing is given by
the eigenvector
\be \label{eq-optimal_dir_squeezing}
\vec{n}_\xi^{(0)}(\tau)=
\vec{n}_-^{(0)} (\tau)= - \sin \theta_{\xi}^{(0)} (\tau) \, \vec{e}_y
+ \cos \theta_{\xi}^{(0)} (\tau) \,
\vec{e}_z
\ee
associated to the eigenvalue $\gamma_-^{(0)}(\tau)$ of the covariant matrix.
One easily finds
\ba
\label{eq:theta_sq}
\theta^{(0)}_{\xi}(\tau)
& = & \arctan \left(
\frac{\gamma_{yz}^{(0)} (\tau)}{ \gamma_+^{(0)} (\tau)-\gamma_{zz}^{(0)} (\tau)}
\right) \nn
\\
&=  &
 \fr{1}{2} \arctan
 \left(
\fr{\bigr\langle \{ \hat{J}_{y} ,  \hat{J}_{z} \} \rangle_\tau^{(0)}}
{\langle  \hat{J}^2_{y} \rangle_\tau^{(0)} -   \langle \hat{J}^2_{z}\rangle_\tau^{(0)}}
 \right)
\ea
where $\{ \cdot, \cdot \}$ denotes the anticommutator.

The direction
of optimization $\vec{n}_{F}^{(0)}$ of the quantum Fisher information
is either given by $\vec{e}_x$
(if $\gamma_x^{(0)} >\gamma^{(0)}_+$) or by the eigenvector
$\vec{n}_+^{(0)}$ associated to the eigenvalue $\gamma^{(0)}_+$
(if $\gamma_x^{(0)} < \gamma^{(0)}_+$). The latter condition is satisfied at times shorter than $t^*$, see Appendix \ref{appD}. As both these eigenvectors
are orthogonal to $\vec{n}_-^{(0)}$ (since
the matrix $\gamma^{(0)}$ is symmetric), it
follows that  coherent spin squeezing and quantum Fisher information
are optimized in perpendicular directions.
At short times, when the state of the system is a squeezed state, this
has a clear physical interpretation: the quantum Fisher information
is maximum in the direction of maximal angular momentum fluctuations, which is
perpendicular to the direction of lowest fluctuations yielding the best squeezing.

\begin{figure}[!ht]
\begin{minipage}{.72\columnwidth}
\includegraphics*[width=\columnwidth]{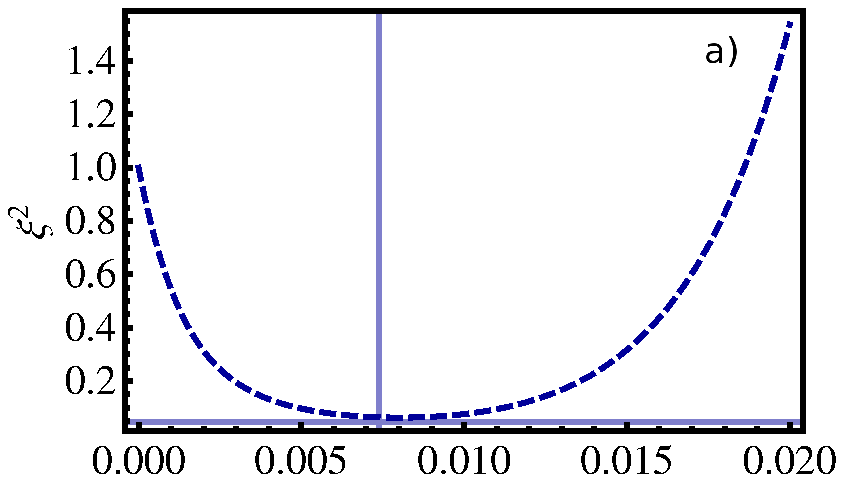}
\end{minipage}
\hfill
\begin{minipage}{.76\columnwidth}
\includegraphics*[width=\columnwidth]{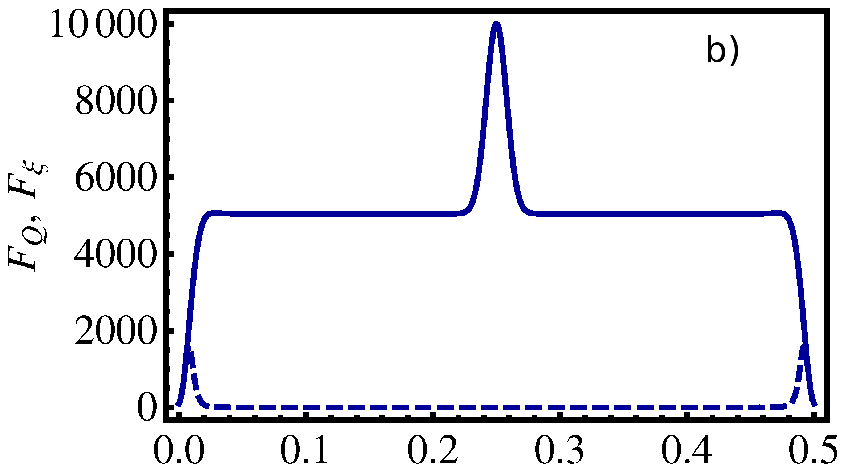}
\end{minipage}
\hfill
\begin{minipage}{.76\columnwidth}
\includegraphics*[width=\columnwidth]{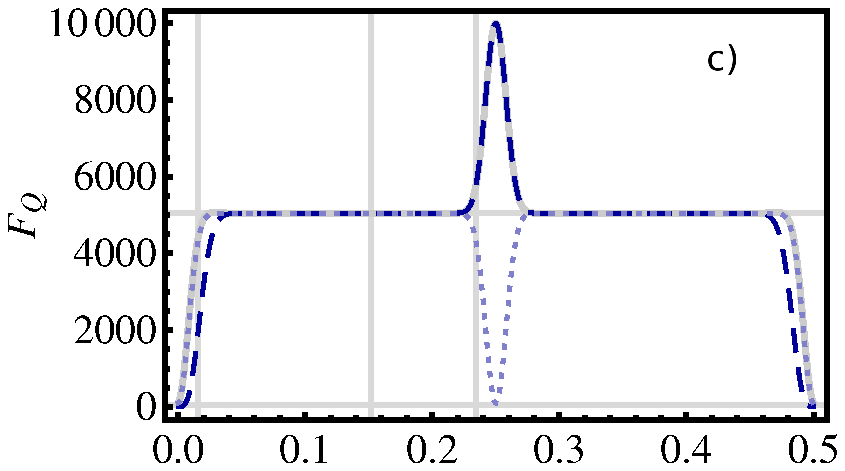}
\end{minipage}
\hfill
\begin{minipage}{.70\columnwidth}
\includegraphics*[width=\columnwidth]{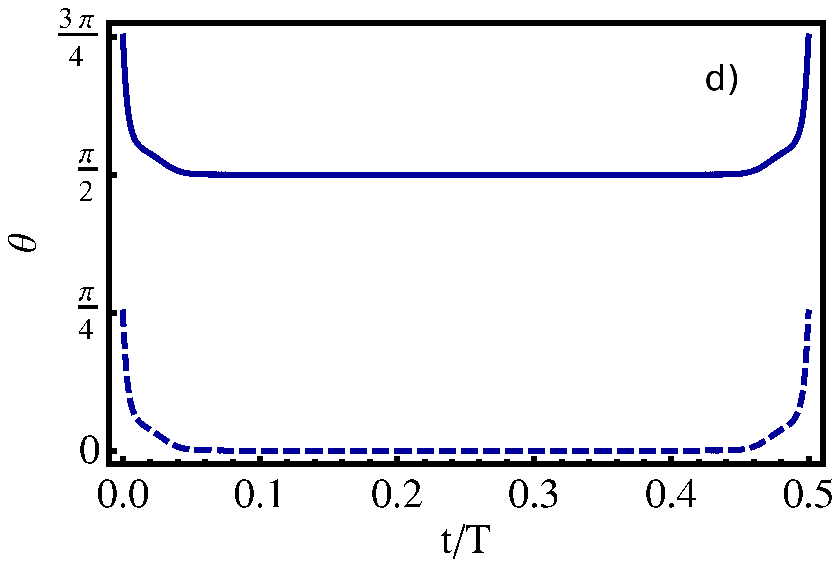}
\end{minipage}
\caption{(Color online) a) Coherent spin squeezing and b) quantum Fisher
information during the quenched dynamics of a BJJ with $N = 100$ atoms
as a function of time
(in units of the revival time  $T$) in the absence of noise.
The dashed line in the second panel represents the parameter $F_{\xi} = N/ \xi^2$.
Horizontal and vertical gridlines in panel a):
minimum of the coherent spin squeezing and corresponding time
$t_{\text{min}}$ (see text).
c) Non-optimized
quantum Fisher information along the $x$-axis (dashed line) and
the $y$-axis (dotted line). For comparison, the
optimum quantum Fisher information of panel b) is also shown (gray solid line).
The vertical gridlines correspond from right to left to the time $t=t_{fs}$ of formation
of the first macroscopic superposition, see Eq.(\ref{eq:time_first_superposition});
to $t=t^*$, see Appendix~\ref{appD};
and to $t=T/4 - t_{fs}$. The horizontal gridline shows the shot-noise level $F_Q = N$.
d) Angles $\theta_{\xi}^{(0)}$ in Eq.(\ref{eq:theta_sq})
(dashed line) and $\theta_{F}^{(0)}$ (solid line)  giving the optimizing direction for the spin squeezing and the quantum Fisher information as a function of time.}
\label{fig:squeezing_fisher_no_noise}
\end{figure}

The Fisher information (\ref{eq-Fisher_info_no_noise}) and
the squeezing
parameter (\ref{eq:squeezing_like_davi2}) obtained from
Eqs.(\ref{eq-eigenvalue_x}) and
(\ref{eq:lambdas_sol_sec})
are shown as a function of time in Fig. \ref{fig:squeezing_fisher_no_noise}.
At short times, the
coherent spin squeezing (a)) is below
one, indicating the presence of a squeezed state.
To compare spin squeezing and Fisher information, we introduce the parameter
$F_\xi=N/\xi^2$. This parameter was shown in \cite{Smerzi09} to coincide  at short times
with the Fisher  information, indicating that $F_Q$ and $\xi$ provide essentially the same
information for squeezed states at such times. This property
is illustrated in Fig.\ref{fig:squeezing_fisher_no_noise}-b), and is demonstrated in the large $N$ limit in Appendix \ref{appE}.
At larger times, the squeezing parameter first reaches a minimum and then
grows to values larger than one (that is,
$F_\xi$ decreases and becomes smaller than $N$). This
 does not imply that the atomic state is not useful for interferometry since, as
described in Sec.\ref{secII}, the squeezing criterion is only a sufficient condition for
 useful entanglement \cite{Smerzi09}. Indeed, the quantum Fisher information
increases above the shot noise level $F_Q=N$ until it reaches a  plateau value $F_Q\simeq N(N+1)/2$  (see Appendix \ref{appE}), at the time
\be
\label{eq:time_first_superposition}
t_{fs} \sim T/\sqrt{N}.
\ee
This time corresponds to the time of formation of the ``first'' (in chronological order)
multicomponent superposition, as one can infer from the following argument. The largest
number of phase states of size $\sqrt{N}/2$ (see Sec.\ref{sec-def-Fisher})
which can be put on the equator of the Bloch sphere of radius $N/2$ is
$q_{\text{max}}\simeq 2 \pi N/\sqrt{N}=2 \pi \sqrt{N}$.
The time of formation of the  multicomponent superposition with the highest number of
phase states is $t_{fs}  = T/(2q_{\text{max}})$,
leading to Eq.~(\ref{eq:time_first_superposition}).
Note that $t_{fs}$ is also the time scale for phase diffusion, that is,
 for the decay of the visibility (\ref{eq-visibiblity}).
It is seen in Fig.\ref{fig:squeezing_fisher_no_noise} that
$F_Q$ diplays a sharp
maximum at $t=t_2=T/4$, in correspondence to the two-component macroscopic superposition
which has the highest possible Fisher information
$F_Q=N^2$. This result is not surprising since this  two-component superposition
(\ref{eq-q-component_cat_state})
is the analog in the phase variable of a NOON state, from which it can be obtained
by a $\pi/2$ rotation around the $y$-axis.
Panel c) of Fig.\ref{fig:squeezing_fisher_no_noise} shows the
Fisher information in the directions $\vec{e}_x$ and $\vec{e}_y$.
In the time regime corresponding to the plateau,
they are almost equal due to the symmetry of the multicomponent superpositions
(this means that the eigenvalues $\gamma_x^{(0)}$ and $\gamma_+^{(0)}$ are almost degenerate).
As one approaches the two-component superposition, the optimizing direction changes to
the $x$-axis, which is the direction of maximal angular momentum fluctuations.

In panel d) of Fig.\ref{fig:squeezing_fisher_no_noise},
the angle  $\theta_{\xi}^{(0)}$ giving the direction of
highest spin squeezing in the $(yOz)$ plane is represented as a function of time together
with the corresponding angle $\theta_{F}^{(0)}$ for the Fisher information, which
gives the optimizing direction $\vec{n}_{F}^{(0)}$ of the Fisher information
according to Eq.(\ref{eq:vector_generic}).
Table II  summarizes the aforementioned results.
Some analytical results obtained for short, intermediate and long times in the limit $N \gg 1$ of a large number of atoms are given in Appendix\ref{appE}.

\begin{widetext}
\begin{table}
\begin{tabular}{|l|c|c|}
\hline
Time &  Optimum quantum Fisher information $F_Q$  & Optimizing direction
\\
\hline
$t = 0$                                &  ${N}$ &  degenerate in $(yOz)$ plane
\\
\hline
$0 \leq t \lesssim T/N$  &  $\nearrow$ \quad ,\quad   given by Eq.(\ref{eq-Fisher-short_time}) &
$ -\cos \theta^{(0)}_{\xi}(t)\, \vec{e}_y - \sin \theta^{(0)}_{\xi}(t) \,\vec{e}_z$
\\
\hline
$T/N \ll  t \leq t_{\text{min}} $   &  $\nearrow$ \quad ,\quad given by Eq.(\ref{eq-Fisher-short_time})   &
$\simeq \vec{e}_y$
\\
\hline
$t_{\text{min}}< t \lesssim t_{fs}$ & $\nearrow$ \quad ,\quad
$3^{1/3} N^{5/3}  <  F_Q \lesssim 0.4323 N^2$ &
$\simeq \vec{e}_y$
\\
\hline
$t_{fs} \ll t \leq t^*$ & $\longrightarrow$ \quad ,\hspace{1cm} $F_Q \simeq N(N+1)/2$ & $\simeq \vec{e}_y$    \\
\hline
$ t^* < t \leq T/4$                & $\nearrow$ \quad ,\quad
$ N(N+1)/2 \lesssim F_Q \leq F_Q(T/4) =N^2$ & $\vec{e}_x$ if $N$ is even, $\simeq \vec{e}_y$ if $N$ is odd   \\
\hline
\hline
Time        &  Optimum coherent spin squeezing parameter $F_{\xi} \equiv N/{\xi^2}$ & Optimizing direction \\
\hline
$t = 0$     & $N$    & degenerate in $(yOz)$ plane
\\
\hline
$0 \leq t \lesssim T/N$  &  $\nearrow$ \quad ,\quad $F_\xi \simeq F_Q$
& $-\sin \theta^{(0)}_{\xi}(t)\, \vec{e}_y + \cos \theta^{(0)}_{\xi}(t) \,\vec{e}_z
$
\\
\hline
$T/N \ll t \leq t_{\text{min}} $  & $\nearrow$ \quad ,\quad
${N} < F_{\xi} \leq  F_{\xi}(t_{\text{min}})= 2N^{5/3} 3^{-2/3}$ &  $\simeq \vec{e}_z$
\\
\hline
$t_{\text{min}}< t \lesssim t_{fs}$ & $\searrow$ \quad ,\quad $N e^{-1/2} \lesssim F_{\xi} <  2 N^{5/3} 3^{-2/3} $  &
$\simeq \vec{e}_z$
\\
\hline
$t_{fs} \ll t \leq t^*$  & $\searrow$ \quad ,\quad $N/3^{N/2-1} \leq  F_{\xi} \ll N $ &
$\simeq \vec{e}_z$
\\
\hline
$ t^* < t \leq T/4$                & $\searrow$ \quad ,\quad  $0 <  F_{\xi}<  N/ 3^{N/2-1}$
 &
$\simeq \vec{e}_z
$     \\
\hline
\end{tabular}
\label{tab:tab2}
\caption{Optimum coherent spin squeezing parameter, optimum quantum Fisher information and corresponding optimizing directions during the quenched dynamics of a Bose Josephson junction
in the absence of noise for $N \gg 1$. The arrows indicate whether the function is increasing or decreasing with time in a given time interval.
}
\end{table}
\end{widetext}

\subsection{Dynamics in the presence of noise}
\label{sse:IV(B)}

Let us now consider the effect of the phase noise introduced in Sec.\ref{sec-noisy_BJJ}  on  the results obtained in the previous subsection.
For coherent spin squeezing the calculation can be carried out analytically.
We start with the observation that even in the presence of noise  $\langle \hat{J}_y\rangle_t = \langle \hat{J}_z \rangle_t=0$ and more generally
the angular-momentum covariance matrix $G$ defined in Eq.(\ref{eq:def_matrix_gamma_0})
has the same structure as
the matrix (\ref{eq:cov_matrix_unitary_t_dep})
in the noiseless case.
Therefore, the arguments used in  Appendix \ref{appC} can be taken over to the noiseful case.
We thus conclude that
the squeezing parameter $\xi^2$ is minimum in the $(yOz)$-plane, and is given by
Eq.(\ref{eq:squeezing_like_davi2}), evaluated for the corresponding quantities in the presence of noise. In particular, the bare visibility  $\nu^{(0)}$, Eq.(\ref{eq-visibiblity}), should be replaced by the  visibility  $\nu$ in the presence of noise ~\cite{Ferrini_10},
\be
\label{eq-visibiity_with_noise}
\nu (t) = \frac{2}{N} \langle \hat{J}_x \rangle_t =
e^{-a^2(t)/2} \nu^{(0)}(t)\,,
\ee
and
$\gamma^{(0)}_-$ by the lowest eigenvalue
$G_-$ of the restriction of the covariance matrix $G$ to the $(yOz)$-plane.

We are now going to compute $G_-$ and the spin squeezing parameter explicitly. In order to do so, 
we need to perform the averages in the presence of noise using the full density matrix $\hat \rho(t)$: $\langle \ldots\rangle_t=\tr(\ldots \hat \rho(t))$. They   are related to those in the absence of noise according to
\be \label{eq-mean_J_with_noise}
\langle \hat{J}_i \rangle_t
= \int_{-\infty}^\infty d \phi \,f(\phi,t)
 \langle e^{i \phi \hat{J}_z} \hat{J}_i e^{-i \phi \hat{J}_z} \rangle_t^{(0)}
\ee
where the expectation value inside the integral is taken for the
pure state $|\psi^{(0)}(t) \rangle$ in the absence of noise.
The rotated angular momentum operators in the above expectation value  are equal to
$\cos \phi \hat{J}_x - \sin \phi \hat{J}_y$,
$\sin \phi \hat{J}_x + \cos \phi \hat{J}_y$, and $\hat{J}_z$ for
$i=x,y$, and $z$, respectively.
A similar derivation holds  for
$\langle \{ \hat{J}_i,\hat{J}_j\} \rangle_t=\tr [ \hat{\rho}(t) \{ \hat{J_i},\hat{J}_j\} ]$, with the result
\ba
\langle \hat{J}_{z}^2\rangle_t
& = &
\langle \hat{J}_{z}^2\rangle_t^{(0)}
= \fr{N}{4} \nn \\
\langle \hat{J}_{y}^2\rangle_t
& = & \frac{1-e^{-2 a^2 (t)}}{2}
 \langle \hat{J}_{x}^2\rangle_t^{(0)}
+
\frac{1+e^{-2 a^2 (t)}}{2} \langle \hat{J}_{y}^2\rangle_t
 \nn
\\
 \langle \{ \hat{J}_{y}, \hat{J}_z \} \rangle_t
& = &
e^{-a^2(t)/2} \langle \{ \hat{J}_{y}, \hat{J}_z \} \rangle_t^{(0)}
\nn
\\
 \langle \{ \hat{J}_{x}, \hat{J}_y \} \rangle_t
& = &  \langle \{ \hat{J}_{x}, \hat{J}_z \} \rangle_t    = 0
\,.
\ea
Finally, the submatrix matrix $G'(t)$ reads
\begin{widetext}
\be \label{eq-covarince_matrix_with_noise}
G' (\tau)
=
\left(
\begin{array}{cc}
\fr{N}{8} \bigl[ - e^{-2a^2(\tau)} (N - 1) \cos^{N-2}\lt 2 \tau \rt + (N +1) \bigr]
&
\hspace*{2mm} \frac{1}{4} e^{-\frac{a^2 (\tau)}{2}} N(N -1)  \cos^{N-2}\lt \tau \rt \sin\lt \tau \rt
\\
\vspace*{0.2cm}
\frac{1}{4} e^{-\frac{a^2 (\tau)}{2}} N(N -1)  \cos^{N-2}\lt \tau \rt \sin\lt \tau \rt
&
\frac{N}{4}
\end{array}
\right)\,.
\ee
Thus, by Eqs.(\ref{eq:squeezing_like_davi2}), (\ref{eq-visibiity_with_noise}) and
(\ref{eq-covarince_matrix_with_noise}), one has
\ba
\label{eq-squeezing_with_noise}
\xi^2(\tau)
& = &
\frac{1}{4{\nu^{(0)}}^2(\tau)}
 \lqu -e^{-a^2 (\tau)} (N -1) \cos^{N-2}\lt 2 \tau \rt
 + e^{a^2 (\tau)}  (N +3) \right.
\nn
\\
& &
\left.
  - (N -1)e^{a^2 (\tau)} \sqrt{(1 - e^{-2 a^2 (\tau)} \cos^{N-2}\lt 2 \tau \rt )^2
   + 16 e^{-a^2(\tau)} \cos^{2(N-2)}\lt \tau \rt \sin^2 \lt \tau \rt} \rqu.
\ea
\end{widetext}
The angle which identifies the optimal squeezing direction is given by
Eq.(\ref{eq:theta_sq}), in which the matrix $\gamma^{(0)'}$ should be replaced
by $G'$.

\begin{figure}
\begin{minipage}{.72\columnwidth}
\includegraphics*[width=\columnwidth]{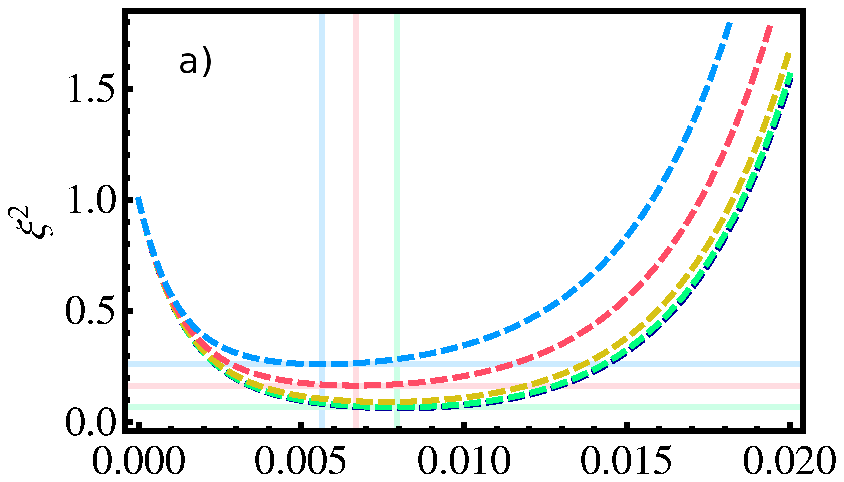}
\end{minipage}
\hfill
\hspace{-1.0cm}
\begin{minipage}{.76\columnwidth}
\includegraphics*[width=\columnwidth]{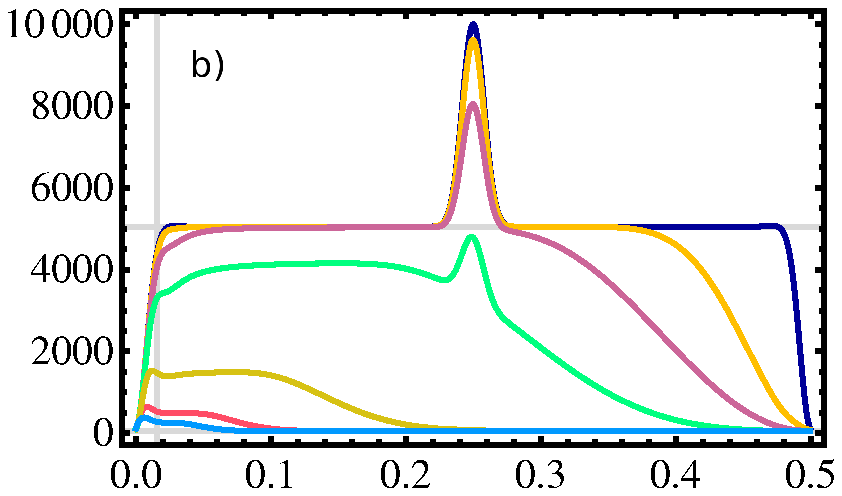}
\end{minipage}
\begin{minipage}{.77\columnwidth}
\includegraphics*[width=\columnwidth]{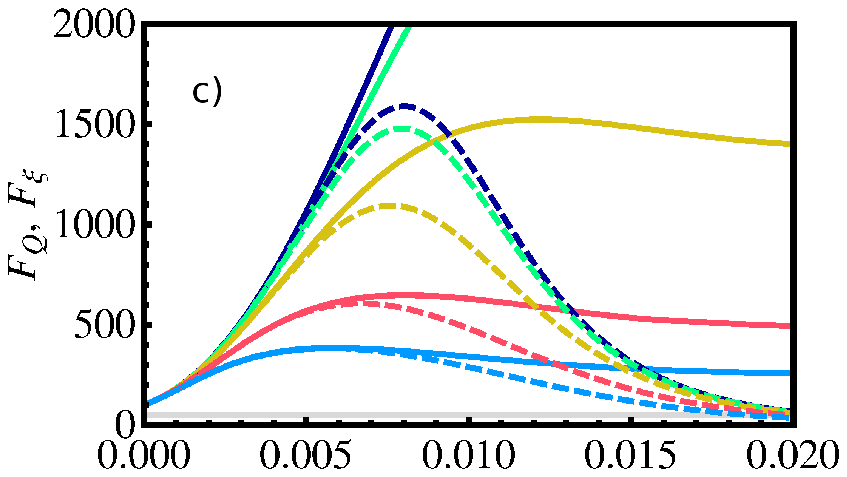}
\end{minipage}
\begin{minipage}{.70\columnwidth}
\includegraphics*[width=\columnwidth]{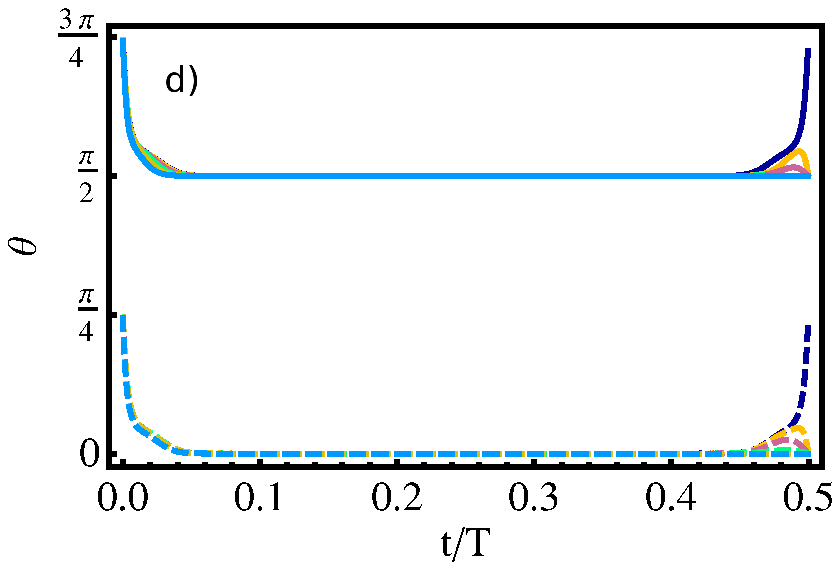}
\end{minipage}
%
\caption{(Color online) Coherent spin squeezing  and quantum Fisher information
in the presence of noise as a function of time in units of $T$ during the quenched dynamics of a
 BJJ. The parameters used are  $N = 100$, $\chi = \pi\,$Hz. a) Spin squeezing
$\xi^2$ for (from top to bottom) $\delta \lambda = 15,10,5$, and $0\,$Hz. Horizontal and vertical
gridlines: minimum of $\xi^2$ and corresponding time $t_{\rm{min}}$.
b) Fisher information $F_Q$ for (from top to bottom)
$\delta \lambda = 0,0.4,1,2,5,10$, and $15\,$Hz; the horizontal and vertical gridlines
correspond to   $F_Q = N(N+1)/2$ and
 $t=t_{fs}=T/\sqrt{N}$. c) Zoom on the quantum
 Fisher information (solid lines) and $F_\xi = N/\xi^2$ (dashed lines) for
 $\delta \lambda = 0,2,5,10$, and $15\,$Hz  (from top to bottom).  d) Angles
$\theta_F$ and $\theta_\xi$ giving the optimizing direction of $F_Q$ (solid lines)
and $\xi^2$ (dashed lines) as a function $t/T$, for the same noise levels.}
\label{fig:squeezing_fisher_noise}
\end{figure}

We proceed by illustrating our results for the squeezing parameter in the presence of phase
noise. For the calculations we have chosen a noise range of direct experimental relevance,
as extracted from the fit of the visibility decay data in Fig.4.15 of Ref.~\cite{Gross_thesis}
with our prediction given by Eq.(\ref{eq-visibiity_with_noise}). For the noise variance
$a^2(\tau)$ we take the
short-time behavior $a^2(\tau) = (\delta \lambda/\chi)^2 \tau^2$
expressed by Eq.(\ref{eq-a2_short_time}) since the experimental visibility
 exhibits a gaussian decay even for small interactions $\chi$ \cite{Gross_thesis}.
This indicates that in the time regime $0 \leq t \lesssim t_{fs}$,
the phase noise in the experiments has strong time correlations (colored noise).
The squeezing parameter as a function of time is shown in Fig.\ref{fig:squeezing_fisher_noise}-a).
  As seen in the figure, the presence of noise degrades the squeezing, as its
 minimum value  increases at increasing noise strength. We also notice that the time for
optimal squeezing $t_{\rm{min}}$ is slighly shorter than in the noiseless case.
As is shown in Appendix \ref{appE} in the limit of a large number of atoms
we find that the minimum value of $\xi^2 (\tau)$ is
\begin{equation}
\label{eq:squeeznoise}
\xi^2_{\rm{min}}\simeq (\xi^{(0)}_{\rm{min}})^2 + N^{-1}(\delta \lambda/\chi)^2
\end{equation}
this minimum being reached to leading order at the same time $\tau_{\rm{min}}=\tau_{\rm{min}}^{(0)}=3^{1/6} N^{-2/3}$ as in the noiseless case.
This means that by increasing the number of atoms
the noise becomes less efficient in limiting the highest squeezing
which can be reached during the dynamical evolution. This results from  the
fact that the time $t_{\rm{min}}$ at which this highest squeezing is produced
is shorter for larger  $N$, whereas the effect of the noise on the
density matrix (\ref{eq:density_matrix_9}) is independent of $N$, as stressed
in Sec.\ref{sec-noisy_BJJ}.
The angle $\theta_\xi(t)$ which identifies the optimizing squeezing direction
is represented in dashed lines at various noise levels in Fig.\ref{fig:squeezing_fisher_noise}-d). As discussed in Appendix \ref{appE}, similar to the noiseless case,
$\theta_{\xi}$ almost vanishes
at intermediate times $N^{-1} \ll \tau \ll\pi/2-N^{-1}$.

The evaluation of the optimum quantum Fisher information (\ref{eq:covariance_general})
requires a numerical diagonalization
of the density matrix $\hat{\rho}(t)$ given by Eq.(\ref{eq:density_matrix_9}).
For the time dependence of
$a^2(t)$ we take again the short-time approximation given in Eq.(\ref{eq-a2_short_time}), even if
 there is no experimental evidence which justifies such a choice at times
$t \sim T$.
This choice corresponds to the worst possible scenario for decoherence, as in the markovian regime the
 dependence of $a^2(t)$ is weaker (see Eq.(\ref{eq-a2_Markov}))~\cite{Ferrini_10}.
 The behaviour of $F_Q$ as a function of time in the presence of
 noise  results from the competition of two phenomena:
(i) in the absence of noise, at short times the quantum Fisher information
 grows from its initial value $F_Q=N$ to the plateau value
$F_Q=N(N+1)/2$ in a time interval $t_{fs}\sim T/\sqrt{N}$ which shrinks
as $N$ becomes larger,
and  (ii) the decoherence exponent $a^2(t)$ is independent of $N$ and
also grows with time.
  As a result,  $F_Q$ reaches a local maximum at a time
 $t_{\rm{max}} \sim t_{sf}$, with a  value which increases with $N$ and decreases
with the noise fluctation $\delta \lambda^2$.

\begin{figure}
\includegraphics*[width=\columnwidth]{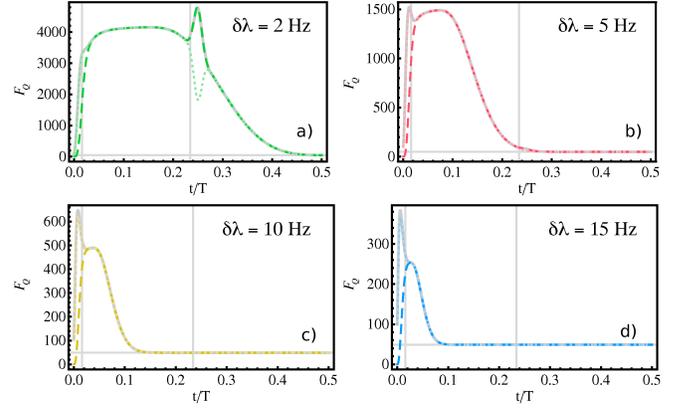}
\caption{(Color online) Direction-dependent quantum Fisher information  in the presence of
 noise as a function of time in units of $T$ during the quenched dynamics of a BJJ with
$N = 100$ atoms and  $\chi = \pi$Hz
for: a) $\delta \lambda = 2$Hz, b) $5$Hz, c) $10$Hz and d) $15$Hz, calculated along the $(Ox)$ direction (dashed lines), the
$(Oy)$ direction (dotted lines) and the optimizing direction (light-gray solid line). After a time $t \sim T/\sqrt{N}$ (left vertical gridlines) the three values are almost the same,
showing that the Fisher information is almost degenerate in the $(xOy)$ plane,
 except around $t = T/4$ if $F_Q$ has a peak at this value
(panel a)). The vertical and horizontal gridlines represent
the times $t=t_{fs}$ and $t=T/4-t_{fs}$ and the value of the
Fisher information
in the limit of large noise intensities given by Eq.(\ref{eq-Fisher_for_rho_infty}).
}
\label{fig:fisher_directions}
\end{figure}

 The quantum Fisher information as a function of time  at various noise levels  is shown
in Fig.~\ref{fig:squeezing_fisher_noise}. The  short-time evolution is similar to the one found for the
noiseless  case, the accumulation of noise correlations being not yet effective.
 In particular, one observes  that $F_Q$ coincides with the
squeezing parameter $F_\xi=N/\xi^2$ at sufficiently small times (panel c).
For
not too large noise intensities, $F_Q$ displays a plateau
at those times which in the noiseless BJJ correspond to the of formation of  macroscopic superpositions.
The value on the plateau
is smaller than in the absence of noise but it is still much above the shot
noise level $F_Q=N$.
This
indicates the presence of useful correlations which remain in spite of the decoherence
effects induced by the noise.  This effect is due to the robustness of the multicomponent superpositions \cite{Ferrini_10} with respect to phase noise 
discussed in Sec. \ref{secV} above.
For
higher noise levels,  the width of the plateau
is reduced and the peak at $t_2\equiv T/4$ corresponding to the two-component superposition in the absence of noise
disappears completely, meaning that decoherence has washed out the
useful quantum correlations at $t_2$
(three bottom curves in the Fig.\ref{fig:squeezing_fisher_noise}-b)).
 In the limit of very large noise intensities the Fisher information at times $t_q$  of formation
of $q$-component superpositions in the noiseless BJJ
 is degenerate in the $(xOy)$ plane and tends to the asympototic value 
\be \label{eq-Fisher_for_rho_infty}
F_Q [ \hat{\rho}_\infty] = \frac{N (N-1)}{2 N + 2}\,,
\ee
which can be readily  obtained from Eqs.(\ref{eq:def_fisher_2}) and (\ref{eq-asymtotic_state}).
As illustrated in Fig.\ref{fig:fisher_directions},  apart from short times
and around the peak at $t_2$,
the optimization direction is in the $(xOy)$-plane
and $F_Q$ is almost degenerate in all directions of this plane,
as in the noiseless case.

As a partial summary, the analysis of the time evolution of the quantum Fisher information
 indicates the build-up of useful quantum correlations at times beyond the spin-squeezing
 regime. In the following we quantify this effect by studying the dependence of $F_Q$
with the noise strength and the particle number.

\begin{figure}
\begin{minipage}{.90\columnwidth}
\includegraphics*[width=\columnwidth]{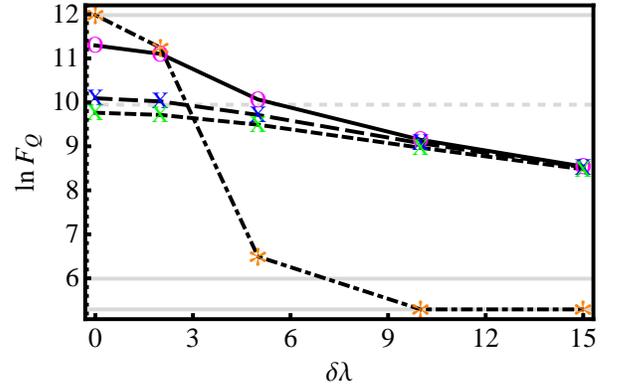}
\end{minipage}
\caption{(Color online) Values of the Fisher information at its local maximum
at time $t_{\rm{max}}$ (solid line, circle markers),  at time $t_2$
(dot-dashed line, star markers)
and at the time $t_{\rm{min}}$ of  maximal squeezing
(long-dashed line, blue cross markers) in a logarithmic scale, as a function of the
energy fluctuation $\delta \lambda$ (in Hz).
For comparison we also plot the squeezing parameter
$F_\xi = N/\xi^2$ at the time $t_{\rm{min}}$
(dashed line, green cross markers) in a logarithmic scale. Gridlines, from top to bottom: Heisenberg limit $N^2$ (solid), approximate value $(2/3^{2/3}) N^{5/3}$ of
$F_\xi(t_{\rm{min}})$ in the absence of noise, see Sec.\ref{sss:IV(A)}
 (dashed), shot noise limit (solid), and limit of $F_Q$
for large noise intensities (solid)  given by Eq.(\ref{eq-Fisher_for_rho_infty}). The parameters used are  $N = 400$ and $\chi = \pi\,$Hz.}
\label{fig:f_vs_noise}
\end{figure}

Figure \ref{fig:f_vs_noise} shows $F_Q(t)$ on a  logarithmic scale,
evaluated at the time $t=t_2 \equiv T/4$ of formation of the two-component
superposition in the noiseless BJJ, as well as the maximum $(F_Q)_{\rm{max}}$
of $F_Q(t)$ in the time interval $0< t <T/8$. This maximum
 corresponds roughly to the value at the plateau
in Fig.\ref{fig:squeezing_fisher_noise}, that is, to the value of $F_Q(t)$
at the times  
of formation of the first multicomponent superpositions.
It can be seen that in the range of noise considered
 $(F_Q)_{\rm{max}}$ stays
above the shot noise level, and is also larger than  the value
$F_Q(t_{\rm{min}})$ at the time
$t_{\rm{min}}$ of  highest squeezing.
The two-component superposition, formed much after the superpositions with a large number of
components,
appears to be too
much degraded by noise to lead to any advantage in interferometry with respect to
separable states. Hence, in this regime multicomponent macroscopic superpositions
provide a convenient alternative to both the squeezed states and the two-component
macroscopic superposition.

\begin{figure}
\begin{minipage}{.98\columnwidth}
\includegraphics*[width=\columnwidth]{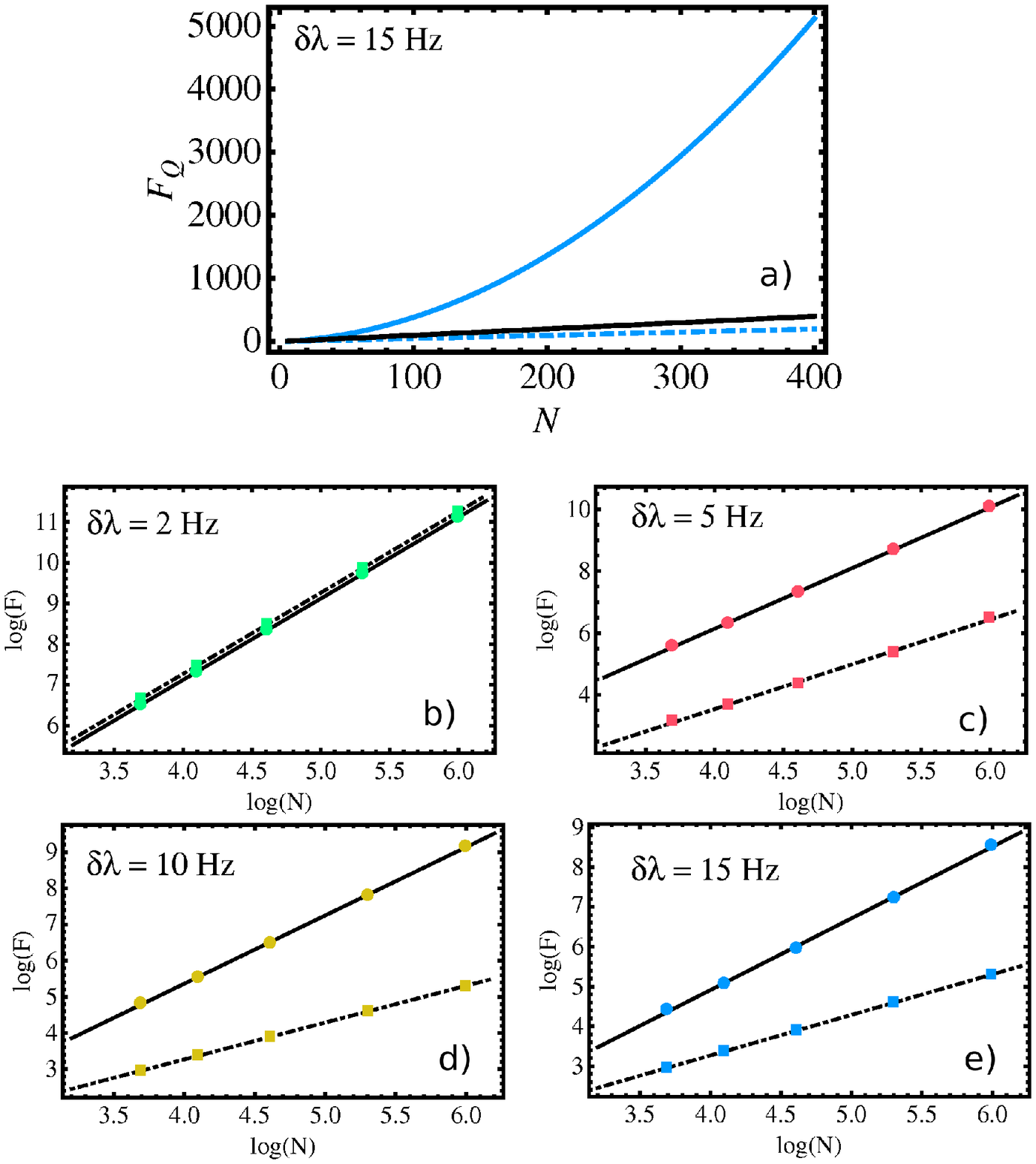}
\end{minipage}
\begin{minipage}{.90\columnwidth}
\includegraphics*[width=\columnwidth]{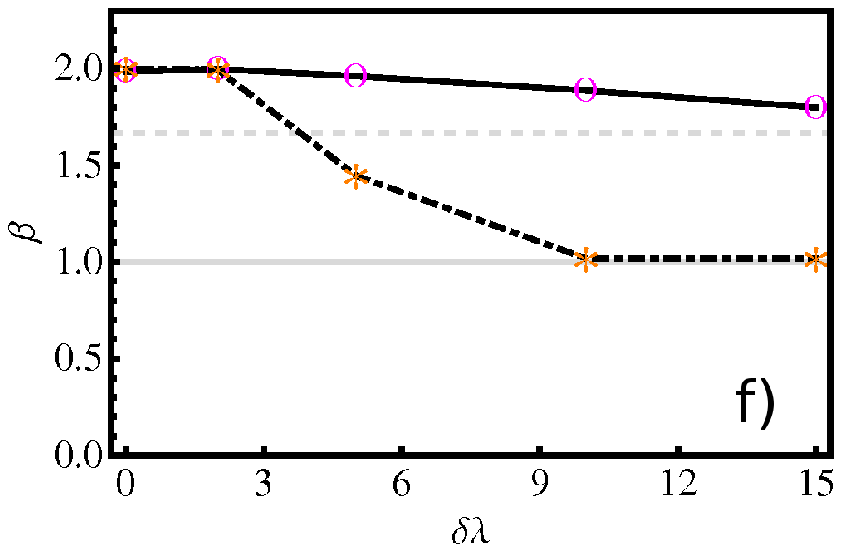}
\end{minipage}
\caption{(Color online) a) Quantum  Fisher information evaluated at the  time of
its local maximum $t_{\rm{max}}$ (blue solid line) and at the time $t_2$ (blue dashed line)
as a function of the number of particles $N$ for $\delta \lambda = 15\,$Hz,
as compared to the shot noise limit (black solid line).
Panels b),c),d),e): same as in a) in a semi-logarithmic scale,
for various noise strengths
$\delta \lambda = 2, 5, 10$, and $15\,$Hz  (from left to right and top to bottom).
f) Exponent $\beta$, extracted by a log-linear fit of the data in a),
as a function of the energy fluctuations $\delta \lambda$ (in Hz) for $t=t_{\rm{max}}$
(solid line, circle markers) and  for $t=t_2$  (dot-dashed line, star markers).
We used $\chi = \pi\,$Hz.}
\label{fig:scaling_fisher}
\end{figure}

We next study  the scaling of the quantum Fisher information  with the particle number, taken at
the time $t_{\rm{max}}$ as before. As it is illustrated in Fig.\ref{fig:scaling_fisher},
at such a time $F_Q$ displays a power-law behaviour $F_Q\sim N^\beta$ with an exponent
$\beta$  depending on the noise strength.
This exponent is extracted from a
log-linear fit of the numerical data, varying $N$ between $50$ and $400$~\cite{note},
 the latter value being
 realistic in the experiments~\cite{Oberthaler10}.
We notice that in the noise range considered
$\beta$ is larger or equal to $5/3$,  which is the
exponent corresponding to the squeezed state at $t=t_{\rm{min}}$
in the absence of noise
(see Sec.\ref{sss:IV(A)})~\cite{nota2}. This confirms the potential
 improvement in interferometry given by the state at time $t_{\rm{max}}$ with respect to the use of
 squeezeed states in the presence of phase noise. For comparison, we also show the
scaling of
 $F_Q$ at the time $t_2$. At that time, $\beta$ decays faster with
 the noise strength, reaching rapidly the shot noise limit $\beta = 1$. This is due to the
 fact that the noise exponent $a^2(t)$ increases with time.


\section{Conclusions}

In this work we have studied the effect of phase noise on the formation of nonclassical
states useful for interferometry created during the quenched dynamics of a Bose Josephson
junction. The knowledge of an exact solution for the dynamical evolution of the state in
the presence of phase noise has allowed us to calculate the quantum Fisher information as a
function of time and its scaling with the particle number at various noise strengths. Due to
the anomalously slow decoherence induced by phase noise on macroscopic superpositions
of phase states, for a realistic choice of noise strengths we have found that multicomponent
 superpositions are more useful for interferometry than squeezed states.
Such superpositions are built
during the dynamical evolution of a noiseless junction at larger times than squeezed states.
The time of formation of the superposition with $\sim \sqrt{N}$ components
depends inversely on the square-root of the total number of
atoms $N$.
When phase noise is affecting the  unitary dynamics of the junction,
these multicomponent superpositions therefore provide an
interesting alternative to the use of the more popular
two-component superposition, which would appear later in a noiseless junction.

\acknowledgements
We thank C. Gross and A. Smerzi for useful discussions, and P. Hyllus for suggesting us
 Ref.\cite{Devi_Sanders}.
We acknowledge financial support from CNRS, the MIDAS project, the Handy-Q project, and  the project ANR-09-BLAN-0098-01.

\appendix
\section{Derivation of the probability distribution $f(\phi,t)$ for gaussian noise }
\label{appA}
The probability distribution $f(\phi,t)$ of the angle
$\phi (t) = - \int_0^{t}{  d\tau \lambda(\tau)}$
is defined as an average over the noise realizations induced by the functional $P[\lambda(t)]$ as in Eq.(\ref{eq:distribution_phase}), or, by Fourier expansion,
\be
f(\phi,t)
 =\fr{1}{2 \pi} \int d P \lqu \lambda(t) \rqu  \int_{- \infty}^{\infty}  d u\,
e^{- i u  \phi(t)} e^{i \phi u}\,.
\ee
We are left to evaluate the Fourier transform of the
average $\overline{ e^{- i u  \phi(t)}} = \int d P [ \lambda(t)] \,e^{- i u  \phi(t)}$.
This is readily done under the hypotesis of a gaussian noise,
\ba
\overline{ e^{- i u  \phi(t)}}&=& \overline{ e^{- i u (\phi(t) - \overline{\phi(t)})}} e^{- i u  \overline{\phi(t)}} \nn \\
&=& e^{- \fr{u^2}{2} \overline{(\phi(t) - \overline{\phi(t)})2}} e^{-i u  \overline{\phi(t)}} \nn \\
&=& e^{- \fr{u^2}{2}  \int_0^{t} d\tau \int_0^{t} d\tau' h(\tau - \tau') } e^{- i u   \overline{\lambda}t} \nn \\
&=& e^{- \fr{u^2}{2} a^2(t)} e^{- i u   \overline{\lambda}t}
\label{eq:expiphi}
\ea
where $h(\tau-\tau') = \overline{\lambda(\tau) \lambda (\tau')} - \overline{\lambda (0)}^2$
is the noise correlation function and
we used that for gaussian variables with $\overline{\xi}= 0$ one has
$\overline{ e^{i u \xi} } = e^{-\fr{u^2}{2} \overline{ \xi^2 } }$.
Using Eq.(\ref{eq:expiphi}) we obtain the expression (\ref{eq:fgauss}) in the main text, according to
\be
 f(\phi,t) = \fr{1}{2 \pi} \int_{- \infty}^{\infty}  du\, e^{- i (\phi +  \overline{\lambda} t) u} e^{- \fr{u^2}{2}  a^2(t)} \,.
\ee

\section{Partial suppression  of phase noise by spin-echo pulses}
\label{appB}

In a recent experiment \cite{Oberthaler10}, phase noise was partially suppressed by  a
spin-echo protocol \cite{Viola_Lloyd_echo}.
Let us assume that the state of interest (for instance, a squeezed state in  \cite{Oberthaler10})
is produced after an evolution time $t$ under the Hamiltonian
(\ref{eq:hamiltonian_noise}).
In the spin-echo protocol,
two
short $\pi$-pulses are send by a laser in resonance with the energies of the two modes
at times $t/2$ and $t$.
The effect of these laser
pulses is to reverse the direction of $\hat{J}_z$, mapped into $-\hat{J}_z$,
in the evolution between $t/2$ and $t$.
Since the noiseless part of the Hamiltonian (\ref{eq:hamiltonian_noise}) is quadratic in
$\hat{J}_z$, it is not affected by the pulses, while the noise part is linear in
$\hat{J}_z$ and is reversed after half of the evolution.
This allows to suppress the effect of the noise if it is strongly correlated
between the two time intervals $[0,t/2]$ and $[t/2,t]$, which appears to be
the case in the experiment of Ref.\cite{Oberthaler10} (see also~\cite{Gross_thesis}).

Our model in Sec.\ref{sec-noisy_BJJ}
can be easily adapted to take into account the residual effect of phase noise
when the spin-echo pulses are applied.
The derivation follows the same lines as in the main text.
Eq.(\ref{eq:density_matrix_9}) still holds provided that we use
$\phi (t) \equiv \int_0^{t}  d\tau \,{\rm Sgn}(\tau-t/2)\lambda(\tau) $, with the sign function
defined as  Sgn$(x)=\pm 1$ for $\pm x>0$. This leads to
\be
a^2_{\text{echo}}(t) =
  \int_0^{t} \!\! d\tau \! \int_0^{t} \!\! d\tau' \, \mathrm{Sgn}(\tau - \fr{t}{2})
 \mathrm{Sgn}(\tau' - \fr{t}{2}) h(\tau - \tau').
\ee
We focus on the short time regime  $t < t_c$. The approximation $h(\tau)\simeq h(0)$ yields no contribution to $a^2_{\text{echo}}(t)$. An expansion to second order is needed,
$h(\tau) = h(0) + \tau h'(0) + \tau^2 h''(0)/2 +O(\tau^3)$. Using the parity
$h (-\tau)=h(\tau)$ of the correlation function (which implies $h' (0)=0$), we
obtain~\cite{rmk_h''(0)<0}
\be
\label{eq:b_t}
a^2_{\text{echo}}(t)
= - \fr{h''(0)}{16}t^4 \,.
\ee
Comparing Eqs. (\ref{eq:b_t}) and (\ref{eq-a2_short_time}),
one sees that the effect of the noise at times $t < t_c$ is considerably reduced
with respect to the case in absence of spin echo.

\section{Demonstration of
Eq.(\ref{eq:squeezing_like_davi2})  for the spin squeezing parameter}
\label{appC}

In the following we show that the spin squeezing parameter $\xi^2(t)$ in a Bose Josephson junction
is always optimized along a direction contained in the $(y O z)$-plane.

Let us observe that
the angular momentum  covariance matrix
$G(t)$ defined by Eq.(\ref{eq:def_matrix_gamma_0}) has vanishing matrix elements $G_{xy}(t)=G_{xz}(t)=0$. In fact, in the absence of noise this matrix
$G(t)=\gamma^{(0)}(t)$ is given by Eq.(\ref{eq:cov_matrix_unitary_t_dep}),
and we have seen in Sec.~\ref{sse:IV(B)} that it preserves the same structure in the presence of
a phase noise.  Thanks to this special structure of $G(t)$,
the fluctuations of the angular momentum operator along an arbitrary
direction $\vec{n}$ given by Eq.(\ref{eq:vector_generic})
is
\ba \label{eq-J_fluctuations}
\Delta {J}_{\vec{n}}(t) &=&
\sum_{i,j=x,y,z} n_i G_{ij}(t) n_j
\\
& = &
\sin^2 \theta \sin^2 \phi \,G_{xx} (t)
 +   \sum_{i,j = y,z} n_i G_{ij}(t) n_j \,.
\nn
\ea
The sum over $i,j$ in the second line can be written
as
$(\sin^2\theta \cos^2 \phi + \cos^2 \theta) \vec{n}'^{T} {G}'(t) \vec{n}'$,
where we introduced the notation
$G'(t)$ for the two-by-two submatrix of $G(t)$
in the plane $(y O z)$ and the normalized vector
\be
\vec{n}' = \fr{n_y \vec{e}_y + n_z\vec{e_z}}
{\sqrt{\sin^2 \theta \cos^2 \phi + \cos^2 \theta}}
\ee
 in this plane. Furthermore, we  observe that during the dynamics of the noisy junction one has
$\langle \hat{J}_y\rangle_t = \langle \hat{J}_z \rangle_t = 0$ at all times.
As a consequence, the expectation values of the angular momentum operators along the
directions defined by Eq.(\ref{eq:vector_perp}) are given by
\ba
\langle \hat{J}_{\vec{p}_1} \rangle_t &=& \cos \phi \langle \hat{J}_x \rangle_t \nn \\
\langle \hat{J}_{\vec{p}_2} \rangle_t &=& - \cos \theta \sin \phi \langle \hat{J}_x
\rangle_t \,.
\ea
%
%
Combining these results  and using the fact that
$G_{xx}(t) \geq 0$, we obtain from Eq.(\ref{eq:squeezing_n})
\ba
\label{eq:spin_squ_opt_final}
&& \hspace*{-2mm}
\fr{N \nu (t)^2}{4} \xi_{\vec{n}}^2(t)
 = \fr{\sin^2 \theta \sin^2 \phi \,G_{xx}(t)}{1 - \sin^2 \phi \sin^2 \theta}
 + \vec{n}'^{T} G'(t) \vec{n}'
\nn
\\
 &&
\geq G_-(t)
=\min_{\vec{n}',\| \vec{n}'\|=1} \lgr \vec{n}'^{T} G' (t) \vec{n}' \rgr
\\ \hspace*{5mm}
& & =   \min_{\vec{n} \in (y O z),\| \vec{n}\|=1} \lgr \vec{n}^{T} \nn G (t) \vec{n} \rgr
 \ea
where $\nu(t)=2 \langle \hat{J}_x\rangle_t/N$ is the visibility and
$G_-(t)$ the smallest eigenvalue of $G'(t)$.
 Since it is clear that the inequality in Eq.(\ref{eq:spin_squ_opt_final}) is an equality
for  $\vec{n}$ equal to the corresponding eigenvector $\vec{n}_-(t)$ of $G_-(t)$,
this demonstrates that the squeezing is minimized along a direction $\vec{n}_-(t)$
contained in the
$(y O z)$-plane.
Combining Eqs.(\ref{eq:squeezing}) and (\ref{eq:spin_squ_opt_final}), we
obtain that the optimum coherent spin squeezing is given by
Eq.(\ref{eq:squeezing_like_davi2}).

\section{Determination of the time $t^*$ when the optimization
direction of the Fisher information changes in the absence of noise}
\label{appD}

If the number  $N$ of atoms is even, the direction of optimization $\vec{n}_F^{(0)}$
of the Fisher information
 in a noiseless Bose Josephson junction
is along $x$-axis at the time $t_2=T/4$ of formation of the
superposition of the two phase states
$|\theta=\pi/2,\phi=0\rangle$ and $|\theta=\pi/2,\phi=\pi\rangle$.
These phase states are indeed diametrically opposite on the equator of Bloch sphere
along this axis.
Since $\vec{n}_F^{(0)}(\tau)=\vec{n}^{(0)}_+ (\tau)$ is in the $(yOz)$-plane at
times $\tau = 2\pi t/T \ll 1$ (see Sec.\ref{sss:IV(A)}),
the optimizing
direction thus changes abruptly from the $(yOz)$-plane to the $x$-axis at some time
$\tau^* \in ]0,\pi/2[$
satisfying
\be \label{eq-implicit_eq_for_t^*}
\gamma^{(0)}_x(\tau^*) = \gamma^{(0)}_{+}(\tau^*)\,.
\ee
In this appendix we determine $\tau^*$  explicitely
 in the limit of large total atom number $N$, supposed to be even.
We may infer from the previous discussion that $\tau^*$ is neither close to
$0$ nor close to $\pi/2$. Consequently, we look for a solution of the implicit
equation (\ref{eq-implicit_eq_for_t^*}) in the interval
$\tau \in [ N^{-\alpha}  , \pi/2 -  N^{-\alpha} ]$, $\alpha$ being a positive exponent
strictly smaller than $1/2$.
Introducing the variables
$u \equiv \cos ( \tau ) \in [0,\cos(N^{-\alpha})]$ and
$v \equiv \cos ( 2 \tau ) \in [-\cos (2N^{-\alpha}),\cos (2N^{-\alpha})]$,
 we obtain with the help of Eqs.(\ref{eq-eigenvalue_x}) and
(\ref{eq:lambdas_sol_sec})
\ba
& & \hspace*{-4mm}
\fr{4 (\gamma^{(0)}_+(\tau) - \gamma^{(0)}_x (\tau))}{N} =
- (N - 1)v^{N - 2} + N u^{2N - 2}
\\
 & & \hspace*{-4mm}
+ 2(N -1) u^{2N - 4} (1 - u^2) + O (N u^{4N-8}) + O ( N v^{2N-4})
\,.
\nn
\ea
Setting $\gamma^{(0)}_+(\tau) = \gamma^{(0)}_x (\tau)$
gives the equation
\be
\label{eq:sol_t_apprx}
\lt 2 - \fr{1}{u^2} \rt^{N - 2} = 2 - u^2 \fr{N-2}{N-1} + O ( e^{-N^{1-2\alpha}})\,.
\ee
For large $N$, the right-hand side of Eq.(\ref{eq:sol_t_apprx}) is strictly larger than one
and is of the order of unity. Hence
the solution must satisfy $|2-u^{-2}|>1$ and  $2-u^{-2} \simeq \pm 1$. We may exclude
the positive sign as
the values $u=\pm 1$ correspond to $\tau \simeq 0$ or $\tau=\pi$ outside the
studied time interval. The relevant solution $u$  of Eq.(\ref{eq:sol_t_apprx})
is thus close to $1/\sqrt{3}$ and smaller than this number.
Let us note that for odd $N$'s, such a solution does not exist; indeed,  in this case
Eq.(\ref{eq-implicit_eq_for_t^*}) has no solution
(see Sec.\ref{sss:IV(A)}).
Let us set $u = 1/(\sqrt{3}(1 + \delta))$.
Then from Eq.(\ref{eq:sol_t_apprx}) we obtain
\be
e^{(N - 2)\ln (1 + 6 \delta + {O}(\delta^2))} = \fr{5}{3} + {O}(\delta) +
{O}\lt \fr{1}{N}\rt
\ee
from which we find
\be
\delta = \fr{1}{6 N}  \ln \lt\fr{5}{3} \rt \lt 1 + {O}\lt \fr{1}{N}\rt \rt
\ee
In terms of the dimensionless time $\tau^*$ we get
\be
\label{eq:time_t_star}
\tau^* = \mathrm{arccos} \lt \fr{1}{\sqrt{3}} \rt +  \fr{\ln (5/3)}{6 \sqrt{2} N} + {O}\lt \fr{1}{N^2}\rt \,.
\ee


\section{Analytical results for the spin-squeezing parameter and the quantum Fisher information in  the large-$N$ limit}
\label{appE}

{\it Short time regime.} At times shorter than the time of formation of the first macroscopic superpositions, ie $0 \leq t \ll t_{fs}$ (or
$\tau \ll 1/\sqrt{N}$), one has
$\gamma^{(0)}_{+}(t) > \gamma^{(0)}_{x}(t)$.
A short-time expansion in Eq.(\ref{eq:lambdas_sol_sec}) followed by the large $N$ limit  yields~\cite{remark-validity_squeezing_formula}
\begin{equation}
\label{eq-Fisher-short_time}
F_Q^{(0)} (\tau) =  4 \gamma^{(0)}_{+}(\tau)
\simeq N
\biggl[ 1 + \biggl( \frac{N^2 \tau^2}{2} + N \tau \sqrt{ 1 + \frac{N^2\tau^2}{4}} \biggr)\biggr].
\end{equation}
%
In this time regime, the visibility (\ref{eq-visibiblity}) is almost equal to one.
In order to compare $F_Q$ with $F_\xi=N/\xi^2$, we determine the ratio
\ba \label{eq-product_xi0_F_Q}
\nn
\frac{F_Q^{(0)}(\tau)}{F_\xi^{(0)} (\tau)}
& \simeq &
\frac{N+1}{2}
- \frac{N-1}{2} \cos^{N-2} ( 2 \tau)
\\
& &
  - (N-1)^2 \cos^{2N-4}( \tau) \sin^2 (\tau)
\ea
by employing Eq.(\ref{eq:squeezing_like_davi2})
and identifying the product $\gamma^{(0)}_{+}(\tau) \gamma^{(0)}_{-}(\tau)$
with the determinant of the
$2 \times 2$ submatrix ${{\gamma}^{(0)}}'(t)$ in Eq.(\ref{eq:cov_matrix_unitary_t_dep}).
At short times  $\tau \ll  N^{-2/3} $, which a posteriori turns out to be the time of optimal squeezing, the RHS of
Eq.(\ref{eq-product_xi0_F_Q})
can be approximated by one, yielding
\be \label{eq-Fisher=squeezing}
F_\xi^{(0)} (\tau) \equiv \frac{N}{\xi^{(0)} (\tau)^2} \simeq   F_Q^{(0)} (\tau )
\ee
as discussed in the main text.
At later times  $\tau \lesssim N^{-2/3}$ the right-hand side
of Eq.(\ref{eq-product_xi0_F_Q})
can be approximated by $1 + N^4 \tau^6/6$. We obtain~\cite{remark-validity_squeezing_formula}
\be \label{eq-squeezing_small_time}
\xi^{(0)} (\tau)^2
 \simeq
  \frac{N \bigl( 1+N^4 \tau^6/6\bigr) }{F_Q^{(0)} (\tau)} \,,
\ee
where $F_Q^{(0)} (\tau)$ is given by Eq.(\ref{eq-Fisher-short_time}) up to a relative correction ${O}(N^{-1/3})$.
The squeezing parameter (\ref{eq-squeezing_small_time}) reaches a minimum
$(\xi_{\rm{min}}^{(0)})^2 \simeq (3/N)^{2/3}/2$ at the time $\tau_{\text{min}}^{(0)}= 3^{1/6}N^{-2/3}$
in the limit $N \gg 1$, as assumed above (see also  Ref.~\cite{Kitagawa93}
where  a different definition of $\xi$ is used,
which however almost coincides with ours at times $t \ll t_{sf}$
because $\nu^{(0)}(t) \simeq 1$).
The value of the Fisher information at $\tau=\tau_{\text{min}}^{(0)}$ is
$F_Q(\tau_{\text{min}}^{(0)}) \simeq 3^{1/3} N^{5/3}$.
The direction of optimization of  $F_Q$
is in the $(y O z)$-plane and is given by the eigenvector
$n_{+}^{(0)} (\tau)$
orthogonal to $n_{-}^{(0)} (\tau)$, that is,
$\phi_{F}^{(0)} (\tau) = 0$ and
$\theta_{F}^{(0)} (\tau) = \theta_{\xi}^{(0)} (\tau)+ \pi/2$.
One finds by using the first equality in (\ref{eq:theta_sq}) that
 $\tan \theta_{\xi}^{(0)} (\tau)
 \simeq ( N \tau /2 + \sqrt{1 + N^2 \tau^2/4})^{-1}$.
The angle  $\theta_{\xi}^{(0)}$ starts from $\pi/4$ at $\tau=0$ and
quickly decreases to the value $0$, to which it is almost equal
at times $\tau \gg N^{-1}$.
At such times $\xi$ and $F_Q$ are optimal along $\vec{e}_z$
and $\vec{e}_y$, respectively (see Eq.(\ref{eq-optimal_dir_squeezing})). These results are summarized in Table II.

{\it Intermediate times.}
In the time regime $\delta t \leq t \leq T/4 - \delta t$ with $\delta t \gg t_{fs}$,
the covariance matrix (\ref{eq:cov_matrix_unitary_t_dep})
 takes the simple following form in the limit $N \gg 1$
\be \label{eq-covariance_plateau_regime}
\gamma^{(0)} (\tau) \simeq
\left(
\begin{array}{cccccccc}
\frac{1}{8} N (N+1)  &  0 &  0
\\
 0 & \frac{1}{8} N (N+1)   &  0 \\
0 &   & \fr{1}{4} N \\
\end{array}
\right)  \, .
\ee
Hence the Fisher information has a plateau  at the value
\be \label{eq-Fisher_on_plateau}
F_Q^{(0)} (\tau)= \frac{N(N+1)}{2}
\ee
whereas the squeezing parameter
\be
F_\xi^{(0)}(\tau) \simeq N {\nu^{(0)}}^2(\tau)
\ee
decreases with time as $\nu^{(0)} (t)$ decreases
(second panel of Fig.\ref{fig:squeezing_fisher_no_noise}).
We have shown in  Appendix~\ref{appC} that if $N$ is even,
the  optimizing direction $\vec{n}_F^{(0)}(\tau)$ of the Fisher information changes  as $\tau$ increases
from the $(yOz)$-plane to the $x$-axis
at the time $\tau^* \simeq \arccos (1/\sqrt{3})$
defined by $\gamma_x^{(0)} (\tau^*)= \gamma_+^{(0)} (\tau^*)$.
Note, however, that any direction in the $(xOy)$-plane gives a Fisher information almost
equal to the optimized value $N(N+1)/2$, as mentioned above
and as it is clear from the structure of the matrix
(\ref{eq-covariance_plateau_regime}).
For an odd number of atoms $N$, the optimal direction $\vec{n}_F^{(0)} (\tau)$ remains in
the $(yOz)$-plane all the way up to $\tau=\pi/2$
(i.e., $\gamma_x^{(0)} (\tau) < \gamma_+^{(0)} (\tau)$ for
any $\tau \in [0,\pi/2]$). More precisely, it is almost along the $y$-axis
(which is the symmetry
axis of the superposition (\ref{eq-q-component_cat_state}) formed at $t=t_2$)
 at times $N^{-1} \ll \tau \leq \pi/2$.

{\it Times $t$ close to $t_2=T/4$.}
At times $t$ such that $|t_2-t |\ll T$, the Fisher information is given by
\be
F_Q^{(0)} (\tau) \simeq \frac{N}{2} \left( N+1 + (N-1) e^{-2 N (\pi/2-\tau)^2} \right)\,.
\ee
It increases monotonously from the plateau value (\ref{eq-Fisher_on_plateau})
at times $t \simeq T/4-t_{sf}$ to the value $N^2$ at the time $t=t_2$ of formation of the
two-component
macroscopic superposition, which has the highest Fisher  information $F_Q=N^2$
allowed by the Heisenberg bound.
The optimal direction of $F_Q$ is along the $x$-axis if $N$ is even and the $y$-axis
if $N$ is odd, and that of $\xi$ is along the $z$-axis in both cases.

{\it Results in the presence of noise.}
The formula generalizing
Eq.(\ref{eq-squeezing_small_time}) for small nonzero noise intensities
$a(\tau) \lesssim N^{-1}$   reads
\be \label{eq-squeezing_with_noise_largeN}
\xi^2(\tau) \simeq \frac{1 + N^4\tau^6/6+ N (\delta \lambda/\chi)^2 \tau^2+{O}(N^{-1/3})}
{1 + \bigl( \frac{N^2 \tau^2}{2} + N \tau \sqrt{ 1 + \frac{N^2\tau^2}{4}} \bigr)}\,,
\ee
where we assumed $\tau \lesssim N^{-2/3}$ and $\delta \lambda/\chi \lesssim N^{1/6}$.
The minimum value of $\xi (\tau)^2$ is given by Eq.(\ref{eq:squeeznoise})
of the main text. The angle $\theta_\xi$ which identifies the optimal squeezing direction satisfies $\tan(\theta_\xi)=e^{-a^2(t)/2}\tan(\theta^{(0)}_\xi)$.



\begin{thebibliography}{33}
\bibitem{nota3} We focus on the effect of phase noise on the preparation of the useful input
state. For studies of the effect of noise during the rotations or the measurements,
see references \cite{Dorner} and \cite{Giovannetti2}.
\bibitem{note} We cannot exclude here that slightly
different values of $\beta$ would appear for larger $N$'s.
\bibitem{nota2} For the chosen interval of noise strengths, the analysis of the scaling
 $F_Q = c N^\beta$
is meaningful as the multiplying constant $c$, which also depends on the noise,
is large enough to
ensure that $F_{Q}(t_{\rm{max}}) \gg N$, as shown in the first panel in Fig.\ref{fig:scaling_fisher}.
\bibitem{rmk_h''(0)<0} $a_{\text{echo}}^2(t)$ is positive since $h''(0)<0$. This follows
from the fact that the correlation function $h$ is of positive type and hence has a
positive Fourier transform.
\end{thebibliography}
\end{document}